\documentclass[12pt]{article}
\usepackage{amsmath}
\usepackage{multirow}
\usepackage{amsfonts}
\usepackage{amssymb,graphics,psfrag}
\usepackage{array,epsfig,multirow,graphicx}

\usepackage{comment}
\usepackage{feynmp}
\usepackage[all,cmtip]{xy}

\numberwithin{equation}{section}
\numberwithin{table}{section}

\usepackage{hyperref}

\def\hybrid{\topmargin -20pt    \oddsidemargin 0pt
        \headheight 0pt \headsep 0pt
        \textwidth 6.25in      
        \textheight 9 in      
        \marginparwidth .875in
        \parskip 5pt plus 1pt
          \jot = 1.5ex
  }
\hybrid

\newcommand{\BPF}{\text{Bl}_3 \mathbb{P}^3_B(D_u,D_v,D_w)}
\newcommand{\BPFs}{\text{Bl}_3 \mathbb{P}^3_B}

\newcommand{\be}{\begin{equation}}
\newcommand{\ee}{\end{equation}} 
\newcommand{\bea}{\begin{eqnarray}}
\newcommand{\eea}{\end{eqnarray}}

\newcommand{\nn}{\nonumber}

\newcommand{\beq}{\begin{equation}}
\newcommand{\eeq}{\end{equation}}

\newcommand{\cC}{\mathcal{C}}

\newcommand{\cS}{\mathcal{S}}

\newcommand{\cref}{{\bf [check ref]}}

\def\blfootnote{\xdef\@thefnmark{}\@footnotetext}
\long\def\symbolfootnote[#1]#2{\begingroup%
\def\thefootnote{\fnsymbol{footnote}}\footnote[#1]{#2}\endgroup}

\begin{document}

\baselineskip=15pt

\begin{titlepage}
\begin{flushright}
\parbox[t]{1.08in}{UPR-1255-T}
\end{flushright}

\begin{center}

\vspace*{ 1.2cm}

{\bf \Large \text{\hspace{-0.6cm}Elliptic Fibrations with Rank Three Mordell-Weil Group:}
F-theory  with U$(1)\times$U$(1)\times$U(1) Gauge Symmetry}

\vskip 1.2cm

\renewcommand{\thefootnote}{}
\begin{center}
 {Mirjam Cveti\v{c}$^{1,2}$,  Denis Klevers$^1$, Hernan Piragua$^1$, Peng Song$^1$}
\end{center}
\vskip .2cm
\renewcommand{\thefootnote}{\arabic{footnote}} 

$\,^1$ {Department of Physics and Astronomy,\\
University of Pennsylvania, Philadelphia, PA 19104-6396, USA} \\[.3cm]

$\,^2$ {Center for Applied Mathematics and Theoretical Physics,\\ University of Maribor, Maribor, Slovenia}\\[.3cm]

{cvetic\ \textsf{at}\ cvetic.hep.upenn.edu, klevers\ \textsf{at}\ sas.upenn.edu,\\ hpiragua\ \textsf{at}\ sas.upenn.edu, songpeng\ \textsf{at}\ sas.upenn.edu} 

 \vspace*{0.8cm}

\end{center}

\vskip 0.2cm
 
\begin{center} {\bf ABSTRACT } \end{center}

We analyze general F-theory compactifications with U(1)xU(1)xU(1) Abelian gauge symmetry by constructing the general elliptically fibered Calabi-Yau manifolds with a rank three Mordell-Weil group of rational sections.  The general elliptic fiber is shown to be a complete intersection of two non-generic quadrics in $\mathbb{P}^3$ and  resolved elliptic fibrations are obtained
by embedding the fiber as the generic Calabi-Yau complete intersection into $\text{Bl}_3\mathbb{P}^3$, the blow-up of $\mathbb{P}^3$ at three points.
For a fixed base $B$, there are finitely many
Calabi-Yau elliptic fibrations. Thus,
F-theory compactifications on these Calabi-Yau manifolds
are shown to be labeled by integral points in reflexive polytopes constructed from the nef-partition of $\text{Bl}_3\mathbb{P}^3$.
We determine all 14 massless matter representations
to six and four dimensions by an explicit study of the codimension two singularities of the elliptic fibration.  We obtain three matter representations charged under all three U(1)-factors, most notably a tri-fundamental representation.
The existence of these representations, which  are not  present in generic perturbative Type II compactifications,  signifies an intriguing universal structure of codimension two singularities of the elliptic fibrations with higher rank Mordell-Weil groups.  We also
compute explicitly the corresponding 14 multiplicities of massless hypermultiplets of a
six-dimensional F-theory compactification for a general base $B$.

\hfill {October, 2013}
\end{titlepage}

\tableofcontents


\newpage
%
\section{Introduction and Summary of Results}

Compactifications of F-theory 
\cite{Vafa:1996xn,Morrison:1996na,Morrison:1996pp} are a very 
interesting and broad class of string vacua, because they are on the one 
hand non-perturbative, but still controllable, and on the other hand realize 
promising particle physics. In particular, F-theory GUTs have drawn a lot 
of attention in the recent years, first in the context of local models following 
\cite{Donagi:2008ca,Beasley:2008dc,Beasley:2008kw,Donagi:2008kj}
and later also in compact Calabi-Yau manifolds 
\cite{Blumenhagen:2009yv,Marsano:2009wr,Chen:2010ts,Grimm:2009yu,Knapp:2011ip},
see e.g.~\cite{Heckman:2010bq,Weigand:2010wm,Maharana:2012tu} for
reviews. Both of these approaches rely on
the well-understood realization of non-Abelian
gauge symmetries that are engineered by constructing codimension 
one singularities of elliptic fibrations
\cite{Vafa:1996xn,Morrison:1996na,Morrison:1996pp,Bershadsky:1996nh} 
that have been classified in 
\cite{kodaira1963compact,tate1975algorithm}.\footnote{A toolbox to construct examples of 
compact Calabi-Yau manifolds with a certain non-Abelian gauge group is 
provided by toric geometry, see
\cite{Candelas:1996su,Candelas:1997eh,Bouchard:2003bu}.} 
In addition, the structure of these codimension one singularities governs
the pattern of matter that is localized at codimension two singularities of the fibration
\cite{Katz:1996xe},  with some subtleties of  higher codimension singularities
uncovered recently in  \cite{Esole:2011sm,Marsano:2011hv,Lawrie:2012gg}.\footnote{For a recent approach based on deformations, 
cf.~\cite{Grassi:2013kha}. See also \cite{Huang:2013yta} for a determination
of BPS-states, including matter states, of (p,q)-strings using the refined 
topological string.}

Abelian gauge symmetries are crucial ingredients for extensions
both of the standard model as well as of GUTs. 
However,
the concrete construction of Abelian gauge symmetries as well as their 
matter content has only recently been addressed systematically in global 
F-theory compactifications. This is due to the fact that U(1) gauge 
symmetries in F-theory are not related to local codimension one 
singularities  but to the global properties of the elliptic fibration of 
the Calabi-Yau manifold. Concretely, the number of U(1)-factors in 
an F-theory  compactification is given by the rank of the Mordell-Weil 
group of the elliptic fibration\footnote{See also 
\cite{Aspinwall:1998xj,Aspinwall:2000kf} for the interpretation of the
torsion subgroup of the Mordell-Weil group as inducing non-simply 
connected non-Abelian group in F-theory.} 
\cite{Morrison:1996na,Morrison:1996pp}, see 
\cite{neron1964modeles,shioda1990mordell,shioda1989,Wazir:2001,silverman2009arithmetic}
for a mathematical background. The
explicit compact Calabi-Yau manifolds with 
rank one \cite{Morrison:2012ei} and the most general rank two 
\cite{Borchmann:2013jwa,Cvetic:2013nia}  
Abelian sector have been constructed recently. In the rank two case,
the general elliptic fiber is the generic elliptic curve in $dP_2$ and 
its Mordell-Weil group is rank two with the two generators induced from
the ambient space $dP_2$. The full six-dimensional spectrum of the 
Calabi-Yau elliptic fibrations with elliptic fiber in $dP_2$ has 
been determined in 
\cite{Cvetic:2013nia,Cvetic:2013jta} and chiral 
compactifications to four dimensions on Calabi-Yau fourfolds with 
$G_4$-flux were constructed in
\cite{Cvetic:2013uta,Borchmann:2013hta}.
We note, that certain aspects of Abelian sectors in F-theory could be 
addressed in local models 
\cite{Donagi:2009ra,Marsano:2009gv,Marsano:2009wr,Dudas:2009hu,Cvetic:2010rq,Dudas:2010zb,Dolan:2011iu,Marsano:2012yc,Mayrhofer:2012zy}.  
In addition, special Calabi-Yau geometries realizing one U(1)-factor 
have been studied in
\cite{Grimm:2010ez,Braun:2011zm,Krause:2011xj,Grimm:2011fx,Cvetic:2012xn,Braun:2013yti}.\footnote{For a systematic study of rational sections on toric K3-surfaces we refer to \cite{Grassi:2012qw}.}

In this work we follow the systematic approach initiated in 
\cite{Morrison:2012ei,Cvetic:2013nia} to construct 
elliptic curves with higher rank Mordell-Weil groups and their resolved 
elliptic fibrations, that aims at a complete classification of all
possible Abelian sectors in
F-theory. We construct the most general F-theory compactifications with 
U$(1)\times$U$(1)\times$U(1) gauge symmetry by building elliptically 
fibered Calabi-Yau manifolds with rank three Mordell-Weil group. Most 
notably, we show that this forces us to leave the regime of 
hypersurfaces to represent these Calabi-Yau manifolds explicitly. In fact, 
the general elliptic fiber in the fully resolved elliptic fibration is 
naturally embedded as the generic Calabi-Yau complete intersection 
into $\text{Bl}_3\mathbb{P}^3$, the blow-up of $\mathbb{P}^3$ at three 
generic points. We show that this is the general elliptic curve 
$\mathcal{E}$ with three rational points and a zero point.  We determine 
the birational map to its Tate and Weierstrass form. All generic Calabi-Yau elliptic fibrations of $\mathcal{E}$ over a given 
base $B$ are completely fixed by the choice of three divisors in the base $B$. Furthermore, we show that 
every such F-theory vacuum corresponds to an integral in certain reflexive polytopes\footnote{The correspondence
between F-theory compactifications and (integral) points in a polytope has been noted
in the toric case \cite{Braun:2013nqa} and in elliptic fibrations with a general rank two Mordell-Weil group \cite{Cvetic:2013uta}.}, 
that we construct explicitly.  

As a next step, we determine the 
representations of massless matter in four- and six-dimensional F-theory 
compactifications by thoroughly analyzing the generic codimension two
singularities of these elliptic Calabi-Yau manifolds. We find 14 different matter representations, cf.~table \ref{tab:MatReps}, 
with various U$(1)^3$-charges.  Note, that the construction leads to representations that are symmetric under permutations of the first two U(1) factors, but not the  third one.
\begin{table}[ht!]
\begin{center}
\begin{tabular}{|c|}
\hline   $\text{U}(1)\times\text{U}(1)\times \text{U}(1)$-charged matter  \rule{0pt}{18pt}  \\[5pt]  \hline  \hline $(1,1,1),$ $(1,1,0),$ $(1,0,1),$ $(0,1,1),$  $(1,0,0),$ $(0,1,0),$ $(0,0,1),$ \rule{0pt}{18pt}  \\[-12pt]  \\[0.1pt]    $(1,1,-1),$ $(-1,-1,-2),$ $(0,1,2),$ $(1,0,2),$  $(-1,0,1),$ $(0,-1,1),$ $(0,0,2)$  \rule{0pt}{18pt}  \\[-7pt]  \\[0.1pt] \hline
\end{tabular}
\caption{Matter representation for F-theory  compactifications with 
a general rank-three Mordell-Weil  group, labeled by their U(1)-charges $(q_1,q_2,q_3)$.}
\label{tab:MatReps}
\end{center}
\end{table}
Interestingly, we obtain three representations charged under 
all three U(1)-factors, most notably a tri-fundamental representation.
Matter in these representations is unexpected in perturbative Type II 
compactifications
and might have interesting 
phenomenological implications.   These results, in particular the 
appearance of a tri-fundamental representation,  indicate  an intriguing 
structure of the codimension two singularities of elliptic fibration with 
rank three Mordell Weil group.

 Furthermore, 
we geometrically derive closed formulas for all 
matter multiplicities of charged hypermultiplets in six dimensions for 
F-theory compactifications on elliptically fibered Calabi-Yau 
threefolds over a general base $B$. As a consistency check, we show that 
the spectrum is anomaly-free.  Technically, the analysis of codimension 
two singularities requires the study of degenerations
of the complete intersection $\mathcal{E}$ in $\text{Bl}_3\mathbb{P}^3$ 
and the computation of the homology classes of the determinantal varieties 
describing certain matter loci.

Along the course of this work we have encountered and advanced a number of 
technical issues. Specifically,  we discovered three birational maps
of the generic elliptic curve $\mathcal{E}$ in $\text{Bl}_3\mathbb{P}^3$ 
to a non-generic form of the elliptic curve of 
\cite{Borchmann:2013jwa,Cvetic:2013nia} in $dP_2$. These maps are 
isomorphisms if the elliptic curve $\mathcal{E}$ does
not degenerate in a particular way. 
The $dP_2$-elliptic curves we obtain are non-generic since one
of the generators of the Mordell-Weil group of 
$\mathcal{E}$, with all its rational points being toric, 
i.e.~induced from the ambient space $\text{Bl}_3\mathbb{P}^3$,  maps to a non-toric rational point. 
It would be interesting to investigate, whether any non-toric rational 
point on $dP_2$ can be mapped to a toric point of $\mathcal{E}$ in 
$\text{Bl}_3\mathbb{P}^3$. In addition, we see directly from this map that the elliptic curve 
in $dP_3$ can be obtained as a special case of the curve $\mathcal{E}$ in 
$\text{Bl}_3\mathbb{P}^3$ .
 
This work is organized as follows. In section \ref{sec:3waysToRk3}
we construct the general elliptic curve $\mathcal{E}$. From the existence
of the three rational points alone, we derive that $\mathcal{E}$ is 
naturally represented as the complete intersection of two non-generic
quadrics in $\mathbb{P}^3$, see section \ref{sec:EasQuadricInt}.
The resolved elliptic curve $\mathcal{E}$ is obtained in section
\ref{sec:EasIntInBlP3} as the generic Calabi-Yau complete intersection
in $\text{Bl}_3\mathbb{P}^3$, where all its rational points are toric, 
i.e.~induced from the ambient space. In section 
\ref{sec:connectionToCubic} we 
construct three canonical maps of this elliptic curve to the non-generic elliptic 
curves in $dP_2$. 
In section \ref{sec:WSF} we find the Weierstrass form of the curve 
$\mathcal{E}$ along with the Weierstrass coordinates of all its rational 
points. We proceed with the construction of  elliptically 
fibered Calabi-Yau manifolds $\hat{X}$ with general elliptic fiber in 
$\text{Bl}_3\mathbb{P}^3$ over a general base $B$ in section 
\ref{sec:EllipticFibrations}.
First, we determine the ambient space and all bundles on $B$ relevant for the
construction of $\hat{X}$ in section 
\ref{sec:ConstructionEllipticFibrations}.  We discuss the basic 
general intersections of $\hat{X}$ in section \ref{sec:BasicGeometry} and
classify all Calabi-Yau fibrations for a given base $B$ in section
\ref{sec:allVacua}. In section \ref{sec:Codim2Sings}  we analyze explicitly the codimension two
singularities of $\hat{X}$, which 
determine the matter representations of F-theory compactifications to
six and four dimensions. We follow a two-step strategy to obtain
the charges and codimension two  loci of the 14 different matter 
representations of $\hat{X}$ in sections 
\ref{sec:singsOfRationalSections} and \ref{sec:SingsWSF}, respectively.  We also 
determine the explicit expressions for the corresponding matter 
multiplicities of charged 
hypermultiplets of a six-dimensional F-theory compactification on a 
threefold $\hat{X}_3$ with general base $B$. Our conclusions and a brief 
outlook can be found in \ref{sec:Conclusions}.
This work contains two appendices:  in  appendix \ref{app:WSFdirectly}  we present explicit formulae for the Weierstrass form of
$\mathcal{E}$,  and in  appendix  \ref{app:nefparts}  we give a short account on nef-partitions, that
have been omitted in the main text.

\section{Three Ways to the Elliptic Curve with Three Rational Points}
\label{sec:3waysToRk3}

In this section we construct explicitly the general elliptic curve 
$\mathcal{E}$ with a rank three Mordell-Weil group of rational points,
denoted $Q$, $R$ and $S$. 

We find three different, but equivalent representations of 
$\mathcal{E}$.
First, in section \ref{sec:EasQuadricInt} we find that $\mathcal{E}$ is 
naturally embedded into $\mathbb{P}^3$ as the complete intersection of 
two non-generic quadrics, i.e.~two homogeneous equations of degree two. 
Equivalently, we embed $\mathcal{E}$ in section \ref{sec:EasIntInBlP3} 
as the generic complete intersection Calabi-Yau into the blow-up 
$\text{Bl}_{3}\mathbb{P}^3$ of 
$\mathbb{P}^3$ at three generic points, which
is effectively described via a nef-partition of the corresponding
3D toric polytope.  In this representation the three rational points
of $\mathcal{E}$ and the zero point $P$ descend from the four 
inequivalent divisors of the ambient space $\text{Bl}_{3}\mathbb{P}^3$.
Thus, the Mordell-Weil group of $\mathcal{E}$ is \textit{toric}.
Finally, we show in section \ref{sec:connectionToCubic} that $\mathcal{E}$ can also 
be represented as 
a non-generic Calabi-Yau hypersurface in $dP_2$. In contrast to the
generic elliptic curve in $dP_2$ that has a rank two Mordell-Weil group 
\cite{Borchmann:2013jwa,Cvetic:2013nia}  which is toric, the onefold in $dP_2$  
we find here exhibits a third rational point, say $S$, and has a rank 
three Mordell-Weil group. This third rational
point, however, is \textit{non-toric} in the presentation of 
$\mathcal{E}$ in $dP_2$. We note that there are three different maps
of the quadric intersection in $\text{Bl}_{3}\mathbb{P}^3$ to an  
elliptic curve in $dP_2$ corresponding to the different morphisms from
$\text{Bl}_{3}\mathbb{P}^3$ to $dP_2$.

We emphasize that in the presentation of $\mathcal{E}$ as a complete 
intersection in $\text{Bl}_3\mathbb{P}^3$ the rank four Mordell-Weil 
group is toric. Thus, as we will demonstrate in section \ref{sec:EllipticFibrations}
this representation is appropriate for the construction of resolved 
elliptic fibrations of $\mathcal{E}$ over a base $B$.

\subsection{The Elliptic Curve as Intersection of Two Quadrics in $\mathbb{P}^3$}
\label{sec:EasQuadricInt}

In this section we derive the embedding of $\mathcal{E}$ with a zero 
point $P$ and the rational points $Q$, $R$ and $S$ into 
$\mathbb{P}^3$ as the intersection of two non-generic quadrics.
We follow the methods described in 
\cite{Morrison:2012ei,Cvetic:2013nia} used for the derivation of 
the general elliptic curves with rank one and two Mordell-Weil groups.

We note that the presence of the four points on $\mathcal{E}$ defines
a degree four line bundle $\mathcal{O}(P+Q+R+S)$ over $\mathcal{E}$. 
Let us first consider a general degree four line bundle $\mathcal{M}$ 
over $\mathcal{E}$. Then the following holds, as we see by employing
the Riemann-Roch theorem:
\begin{itemize}
\item[1.] $H^0(\mathcal{E},\mathcal{M})$ is generated by four sections, that we denote 
by $u',\,v',\,w',\,t'$.
\item[2.] $H^0(\mathcal{E},\mathcal{M}^2)$ is generated by eight sections. 
However we know ten sections of $M^2$, the quadratic monomials in 
$[u':v':w':t']$, i.e.~$u'^2$, $v'^2$, $w'^2$, $t'^2$, $u'v'$, $u'w'$, $u't'$, $v'w'$, 
$v't'$, $w't'$.
\end{itemize}
The above first bullet point shows that $[u':v':w':t']$ are of equal 
weight one and can be viewed as homogeneous coordinates on 
$\mathbb{P}^3$. The second bullet point implies that  
$H^0(2\mathcal{M})$ is generated by sections we already know and
that there have to be two relations between the ten quadratic
monomials in $[u':v':w':t']$, that we write as
\bea\label{eq:1stold}
&\!\!\!\!s_1t'^2+s_2u'^2+s_3v'^2+s_4w'^2+s_5t'u'+s_6u'v'+s_7u'w'+s_8v'w'=s_9v't'+s_{10}w't'\,,&\\
&\!\!\!\!s_{11}t'^2+s_{12}u'^2+s_{13}v'^2+s_{14}w'^2+s_{15}u't'+s_{16}u'v'+s_{17}u'w'+
s_{18}v'w'=s_{19}v't'+s_{20}w't'\,,&\nn
\eea

Now specialize to $\mathcal{M}=\mathcal{O}(P+Q+R+S)$ and assume $u'$ to 
vanish at all points $P,Q,R,S$. By inserting $u'=0$ into 
\eqref{eq:1stold} we should then get four rational solutions 
corresponding to the four points, i.e.~other words \eqref{eq:1stold} 
should factorize accordingly. However, this is not true for generic 
$s_i$ taking values e.g.~in the ring of functions of the base $B$ of an
elliptic fibration\footnote{In contrast, if we were considering an elliptic curve over an algebraically 
closed field, we could set some $s_i=0$  by using the 
$\mathbb{P}GL(4)$ symmetries of $\mathbb{P}^3$ to eliminate some 
coefficients $s_i$. For example, $s_3=0$ can be achieved by making the 
transformation
\begin{equation}
u'\mapsto u'+kv'\,,\qquad \text{with $k$ obeying}\qquad (s_2k^2+s_6k+s_3)=0\,.
\end{equation}
Solving this quadratic equation in $k$ will, however, involve the 
square roots of $s_i$, which is only defined in an algebraically closed 
field. In particular, when considering elliptic fibrations the 
coefficients $s_i$ will be represented by polynomials, of which a 
square root is not defined globally.}
Thus, we have to set the following coefficients $s_i$ to zero,
\beq \label{eq:s_i0}
  s_1=s_3=s_4=s_{11}=s_{13}=s_{14}=0\,.
\eeq  
As we see below in section \ref{sec:EasIntInBlP3}, this can be achieved globally, by blowing up $\mathbb{P}^3$ 
at three generic points.

For the moment, let us assume that \eqref{eq:s_i0} holds and determine 
$P,Q,R,S$. First we note that the presentation \eqref{eq:1stold} for 
the elliptic curve $\mathcal{E}$ now reads
\bea\label{eq:1st}
s_2u'^2+s_5u't'+s_{6}u'v'+s_7u'w'&=&s_9v't'+s_{10}w't'-s_8v'w'\,,\\
s_{12}u'^2+s_{15}u't'+s_{16}u'v'+s_{17}u'w'&=&s_{19}v't'+s_{20}w't'-s_{18}v'w'\,,\nn
\eea
which is an intersection of two \textit{non-generic} quadrics in 
$\mathbb{P}^3$.
Setting $u'=0$ we obtain 
\begin{equation}\label{eq:new1}
0=s_9v't'+s_{10}w't'-s_8v'w'\,,\qquad 0=s_{19}v't'+s_{20}w't'-s_{18}v'w'\,,
\end{equation}
which has in the coordinates $[u':v':w':t']$ the four solutions 
\bea \label{eq:P3coordsPQRS}
&P=[0:0:0:1]\,, \quad Q=[0:1:0:0]\,, \quad R=[0:0:1:0]\,,&\nn\\ \quad& S
= [0:|M^S_1||M^S_3|:-|M^S_1||M^S_2|:-|M^S_3||M^S_2|]\,.&
\eea 
Here we introduced the
determinants $|M^S_{i}|$ of all three $2\times 2$-minors $M_i^S$ 
reading
\beq \label{eq:MSDets}
|M^S_1|=s_9s_{20}-s_{10}s_{19}\,,
\qquad |M^S_2|=s_8s_{19}-s_9s_{18}\,,\qquad 
|M^S_3|=s_8s_{20}-s_{10}s_{18}\,,
\eeq
that are obtained by deleting the $(4-i)$-th column in the matrix
\beq \label{eq:MS}
	M^S=\begin{pmatrix}
		s_9 & s_{10} & -s_8\\
		s_{19}& s_{20} & -s_{18}
	\end{pmatrix}\,,
\eeq 
where $M^S$ is the matrix of coefficients in \eqref{eq:new1}.

It is important to realize that the coordinates
of the rational point $S$ are products of determinants in 
\eqref{eq:MSDets}, in particular when studying elliptic fibrations at 
higher codimension in the base $B$, cf.~section \ref{sec:Codim2Sings}. 
On the one hand, the vanishing loci of the determinant of 
a single determinant $|M^S_i|$ with $i=1,2,3$ indicates the collisions 
of $S$ with $P$, $Q$ and $R$, respectively, i.e.~
\beq
	|M^S_1|=0\,:\,\,\, S=P\,,\qquad |M^S_2|=0\,: \,\,\, S=Q\,,\qquad |M^S_3|=0\,:\,\,\, S=R\,.
\eeq
On the other hand the simultaneous vanishing of all $|M^S_i|$ is 
equivalent to the two constraints in \eqref{eq:1st} getting linearly 
dependent. Then, the elliptic curve $\mathcal{E}$ degenerates to an
$I_2$-curve, i.e.~two $\mathbb{P}^1$'s intersecting at two points, 
see the discussion around \eqref{eq:DegenerationOfQuadrics}, with the 
point $S$ becoming the entire 
$\mathbb{P}^1=\{u=s_9v't'+s_{10}w't'-s_8v'w'=0\}$\footnote{This curve 
can be seen to define a $\mathbb{P}^1$ either using 
adjunction or employing the Segre embedding of $\mathbb{P}^1\times 
\mathbb{P}^1$ into $\mathbb{P}^3$.}. We note that this behavior of $S$ 
indicates that in an elliptic fibration the point $S$ 
will only give rise to a rational, not a holomorphic section of the 
fibration.
 
In summary, we have found that the general elliptic 
curve $\mathcal{E}$ with three rational points $Q$, $R$, $S$ and a zero 
point $P$ is embedded  into $\mathbb{P}^3$ as the intersection of the 
two non-generic quadrics \eqref{eq:1st}. 

\subsection{Resolved Elliptic Curve as Complete Intersection in $\text{Bl}_{3}\mathbb{P}^3$}
\label{sec:EasIntInBlP3}

In this section we represent the elliptic curve $\mathcal{E}$ with a 
rank three Mordell-Weil group as a \textit{generic} complete 
intersection Calabi-Yau in the ambient space $\text{Bl}_3\mathbb{P}^3$.
As we demonstrate here, the three blow-ups in $\text{Bl}_3\mathbb{P}^3$ 
remove globally the coefficients in \eqref{eq:s_i0}. In addition, the 
three blow-ups resolve all singularities of $\mathcal{E}$, that can 
appear in elliptic fibrations. Finally, we emphasize that the elliptic
curve $\mathcal{E}$ is a complete intersection  associated to the 
nef-partition of the polytope of $\text{Bl}_3\mathbb{P}^3$, where we 
refer to appendix \ref{app:nefparts} for more details on 
nef-partitions.

First, we recall the polytope of $\mathbb{P}^3$ and its nef-partition 
describing a complete intersection of quadrics. The polytope $\nabla_{\mathbb{P}^3}$ 
of $\mathbb{P}^3$ is the convex 
hull $\nabla_{\mathbb{P}^3}=\langle\rho_1,\rho_2,\rho_3,\rho_4\rangle$ of the four vertices 
\beq \label{eq:vertsP3}
\rho_1=(-1,-1,-1)\,,\qquad \rho_2=(1,0,0)\,,\qquad \rho_3=(0,1,0)\,,
\qquad \rho_4=(0,0,1)\,,
\eeq
corresponding to the homogeneous 
coordinates $u'$, $v'$, $w'$ and $t'$, respectively.
The anticanonical bundle of $\mathbb{P}^3$ is 
$K^{-1}_{\mathbb{P}^3}=\mathcal{O}(4H)$, where $H$ denotes the 
hyperplane class of $\mathbb{P}^3$.  
Two generic degree two polynomials in the class $\mathcal{O}(2H)$ 
are obtained via \eqref{eq:sectionNEFpart} from the 
nef-partition of the polytope of $\mathbb{P}^3$ into $\nabla_1$, $\nabla_2$  reading
\begin{equation}
\nabla_{\mathbb{P}^3}=\langle\nabla_1\cup \nabla_2\rangle\,,\qquad \nabla_1=\langle\rho_1, \rho_2\rangle\,,\quad \nabla_2=\langle\rho_3, \rho_4\rangle\,,
\end{equation}
where $\cup$ denotes the union of sets of a vector space.
This complete intersection defines the elliptic curve in 
\eqref{eq:1stold} with only the origin $P$.

Next, we describe the elliptic curve $\mathcal{E}$ as a generic 
complete intersection associated to  a nef-partition of 
$\text{Bl}_3\mathbb{P}^3$, the blow-up of  $\mathbb{P}^3$ at three 
generic points, that we choose
to be $P$, $Q$ and $R$ in \eqref{eq:P3coordsPQRS}. We first perform
these blow-ups and determine the proper transform of $\mathcal{E}$ by 
hand, before we employ toric techniques and nef-paritions.

The blow-up from $\mathbb{P}^3$ to $\text{Bl}_3\mathbb{P}^3$ is 
characterized by the blow-down map
\beq \label{eq:BlowDownMap}
 u' = e_1 e_2 e_3 u \,, \qquad v' = e_2 e_3 v \,, \qquad w' = e_1 e_3 w \,, \qquad t' = e_1 e_2 t \,.\qquad
\eeq
It maps the coordinates $[u:v:w:t:e_1:e_2:e_3]$ on 
$\text{Bl}_3\mathbb{P}^3$ to the coordinates on $[u:v:w:t]$ on 
$\mathbb{P}^3$. Here the $e_i=0$, $i=1,2,3$, are the exceptional 
divisors $E_i$ of the the blow-ups at the points $Q$, $R$ and $P$, 
respectively. 
We summarize the divisor classes of all homogeneous coordinates on 
$\text{Bl}_3\mathbb{P}^3$ together with the corresponding $\mathbb{C}^*$-actions
that follow immediately from \eqref{eq:BlowDownMap} as
\beq \label{eq:divclassBlP3}
	\begin{array}{c|c|rrrr}
	
	 & \text{divisor class}&\multicolumn{4}{c}{\mathbb{C}^*\text{-actions}}\\
	\hline
		 u& H-E_1-E_2-E_3\rule{0pt}{1Em}& 1& 1&1 &1\\ 
		 v& H-E_2-E_3&1&0&1&1\\ 
		 w& H-E_1-E_3&1&1&0&1\\
		 t & H-E_1-E_2&1&1&1&0\\
		 e_1& E_1&0&-1&0&0\\
		  e_2& E_2&0&0&-1&0\\ 
		  e_3 & E_3&0&0&0&-1\\
	\end{array}\,
\eeq
Here $H$ denotes the pullback of the hyperplane class $H$ on 
$\mathbb{P}^3$. 
The coordinates $[u:w:t]$, $[u:v:t]$ and $[u:v:w]$ are
the homogeneous coordinates on each $E_i\cong \mathbb{P}^2$, 
respectively, and can not vanish simultaneously. Together with the 
pullback of the Stanley-Reissner ideal of
$\mathbb{P}^3$ this implies the following Stanley Reisner ideal on 
$\text{Bl}_3\mathbb{P}^3$,
\beq \label{eq:SRBlP3}
SR=\{u v t, u w t , u v w, e_1 v,  e_2 w, e_3 t, e_1 e_2, e_2 e_3, e_1 e_3 \}\,.
\eeq
This implies the following intersections of the four independent divisors on $\text{Bl}_3\mathbb{P}^3$,
\beq \label{eq:intsBlP3}
	H^3=E_i^3=1\,,\qquad E_i\cdot H=E_i\cdot E_j=0\,,\quad i\neq j\,.	
\eeq

The proper transform under the map \eqref{eq:BlowDownMap} 
of the constraints \eqref{eq:1st} describing $\mathcal{E}$ read
\bea \label{eq:EonBlP3}
&\!\!\!\!\!\!p_1:=s_2e_1 e_2 e_3 u^2+s_5e_1 e_2  ut+s_{6}e_2 e_3 u v +s_7e_1 e_3 u w  -s_9e_2  vt -s_{10}e_1  wt + 
s_8e_3 v w  \,, & \\
&p_2:=s_{12}e_1 e_2 e_3 u ^2 + s_{15}e_1 e_2  ut + s_{16}e_2 e_3 u v + s_{17}e_1 e_3 u w    - s_{19}e_2  vt - s_{20}e_1  wt+ s_{18}e_3 v w\,. & \nn
\eea
We immediately  see that this complete intersection defines a 
Calabi-Yau onefold in $\text{Bl}_3\mathbb{P}^3$ employing 
\eqref{eq:divclassBlP3}, adjunction and noting that the anti-canonical 
bundle of $\text{Bl}_3\mathbb{P}^3$ reads
\beq
	K_{\text{Bl}_3\mathbb{P}^3}=\mathcal{O}(4H-2E_1-2E_2-2E_3)\,.
\eeq

From \eqref{eq:P3coordsPQRS}, \eqref{eq:BlowDownMap} and 
\eqref{eq:EonBlP3} we readily obtain
the points in $P$, $Q$, $R$ and $S$ on 
$\text{Bl}_3\mathbb{P}^3$. They are given by the intersection of 
\eqref{eq:EonBlP3} with the four inequivalent toric divisors on 
$\text{Bl}_3\mathbb{P}^3$, the divisor $D_u:=\{u=0\}$ and the 
exceptional divisors $E_i$. Their coordinates read
\bea \label{eq:resolvedPoints}
E_3\cap \mathcal{E}\!\!\!&:&\,\,P=[s_{10} s_{19}-s_{20}s_{9}:s_{10} s_{15}-s_{20} s_5:s_{19} s_5-s_{15} s_9: 1:1:1:0]\,, \nn \\
E_1\cap \mathcal{E}\!\!\!&:&\,\,Q=[s_{19} s_{8}-s_{18}s_{9}:1:-s_{19} s_6+s_{16} s_9 :-s_{18} s_6+s_{16} s_8:0:1:1]\,, \nn \\
E_2\cap \mathcal{E}\!\!\!&:&\,\,R=[s_{10} s_{18}-s_{20}s_{8}:-s_{10} s_{17}+s_{20} s_7:1:s_{18}s_7-s_{17}s_8 :1:0:1]\,, \nn \\
D_u\cap \mathcal{E}\!\!\!&:&\,\,S=[0:1:1:1 :s_{19} s_{8} -  s_{18} s_9: s_{10} s_{18} -  s_{20} s_8:s_{10} s_{19} - s_{20} s_9]\,. \nn \\
\eea
Here we made use of the Stanley-Reissner ideal \eqref{eq:SRBlP3} to set 
the coordinates to one that can not vanish simultaneously with $u=0$,
respectively, $e_i=0$. 

We emphasize that the coordinates \eqref{eq:resolvedPoints} are again
given by determinants of $2\times 2$-minors. Indeed, we can write
\eqref{eq:resolvedPoints} as 
\bea \label{eq:PQRSMinors}
	&P=[-|M^P_3|:|M^P_2|:-|M^P_1| :1:1:1:0]\,,\,\, \quad \!\!
Q=[-|M^Q_3|:1:|M^Q_2|:-|M^Q_1|:0:1:1]\,, &\nn \\
&R=[|M^R_3|:-|M^R_2|:1 :|M^R_1|:1:0:1]\,, \quad
S=[0:1:1 :1:-|M^Q_3|: |M^R_3|:-|M^P_3|]\,\,\,\,\,\,\,\,\,&
\eea
Here we defined the matrices  
\beq \label{eq:MatsPQR}
	M^P=\begin{pmatrix}
	-s_{5}  &  s_9    & s_{10}\\
	-s_{15} &  s_{19} & s_{20}
	\end{pmatrix}\,,\quad  M^Q=\begin{pmatrix}
	-s_{6}  &  -s_8    & s_{9}\\
	-s_{16} &  -s_{18} & s_{19}
	\end{pmatrix}\,,\quad  M^R=\begin{pmatrix}
	-s_{7}  &  -s_8    & s_{10}\\
	-s_{17} &  -s_{18} & s_{20}
	\end{pmatrix}
\eeq
with their $2\times 2$-minors $M^{P,Q,R}_i$ defined by deleting the $(4-i)$-th column. 
We emphasize that the minors of the matrix $M^S$ in \eqref{eq:MSDets} can be expressed by
the minors of the matrices in \eqref{eq:MatsPQR} and, thus, $M^S$ does not appear in 
\eqref{eq:PQRSMinors}.
The matrices
$M^{P,Q,R}$ describe the two linear equations that we obtain by setting
$e_3=0$, $e_2=0$ and $e_1=0$ in \eqref{eq:EonBlP3}, respectively. 

It is important to realize that the points $P$, $Q$ and $R$ are always distinct,
as can be seen from \eqref{eq:PQRSMinors} and the Stanley-Reissner ideal
\eqref{eq:SRBlP3} since the exceptional divisors do not mutually intersect. However,
the point $S$ can agree with all other points, if the appropriate minors in \eqref{eq:PQRSMinors}
vanish. In fact, we see the following pattern,
\beq \label{eq:intsSPQR}
	|M^P_3|=0\,:\,\,\, S=P\,,\qquad |M^Q_3|=0\,: \,\,\, S=Q\,,\qquad |M^R_3|=0\,:\,\,\, S=R\,,
\eeq
which will be relevant to keep in mind for the study of elliptic fibrations.

We note that the 
elliptic curve  $\mathcal{E}$ degenerates into an $I_2$-curve if, as 
explained before below \eqref{eq:MS}, the rank of one of the matrices 
in \eqref{eq:MS} and \eqref{eq:MatsPQR} is one\footnote{We emphasize 
that the complete intersection \eqref{eq:1st} 
in $\mathbb{P}^3$ degenerates into only one $\mathbb{P}^1$ and becomes
singular if one matrices in \eqref{eq:MatsPQR} has rank one, in 
contrast to the smooth $I_2$-curve obtained from \eqref{eq:EonBlP3}.}. 
In addition, one particular intersection in \eqref{eq:resolvedPoints} 
no longer yields a point in $\mathcal{E}$, but an entire 
$\mathbb{P}^1$. As discussed below in section \ref{sec:Codim2Sings} the 
points on $\mathcal{E}$, thus, will only lift to rational sections of 
an elliptic fibration of $\mathcal{E}$.

\begin{figure}[ht]
\centering{\includegraphics[scale=0.5]{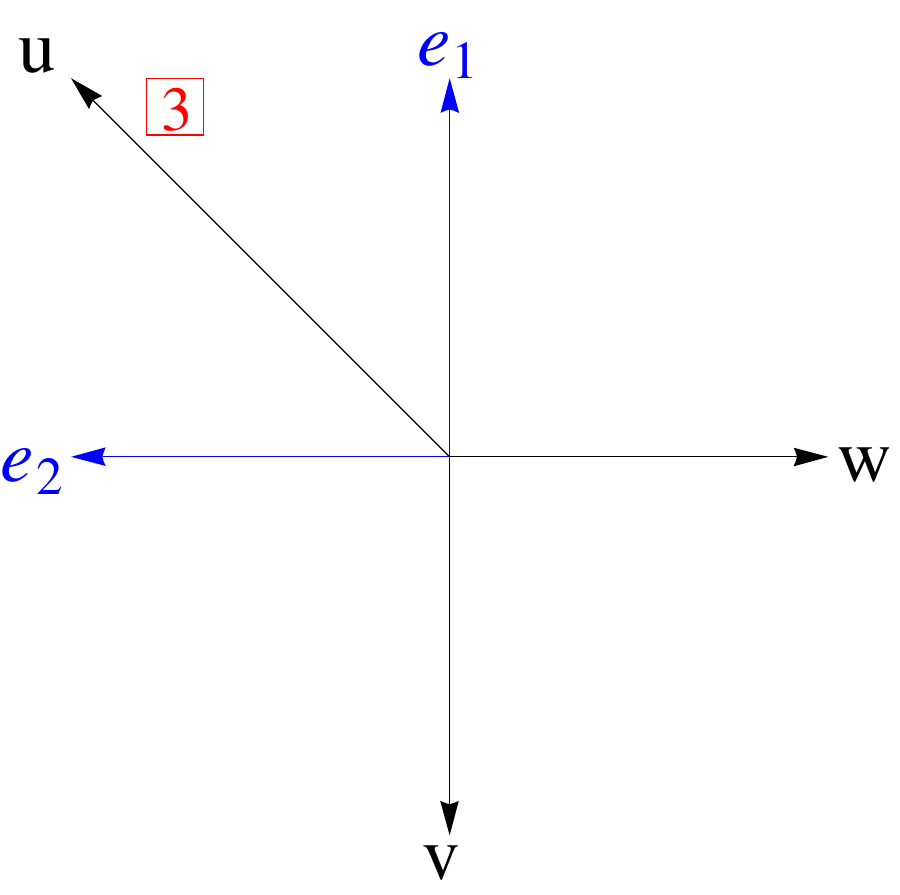}}\\
\includegraphics[scale=0.5]{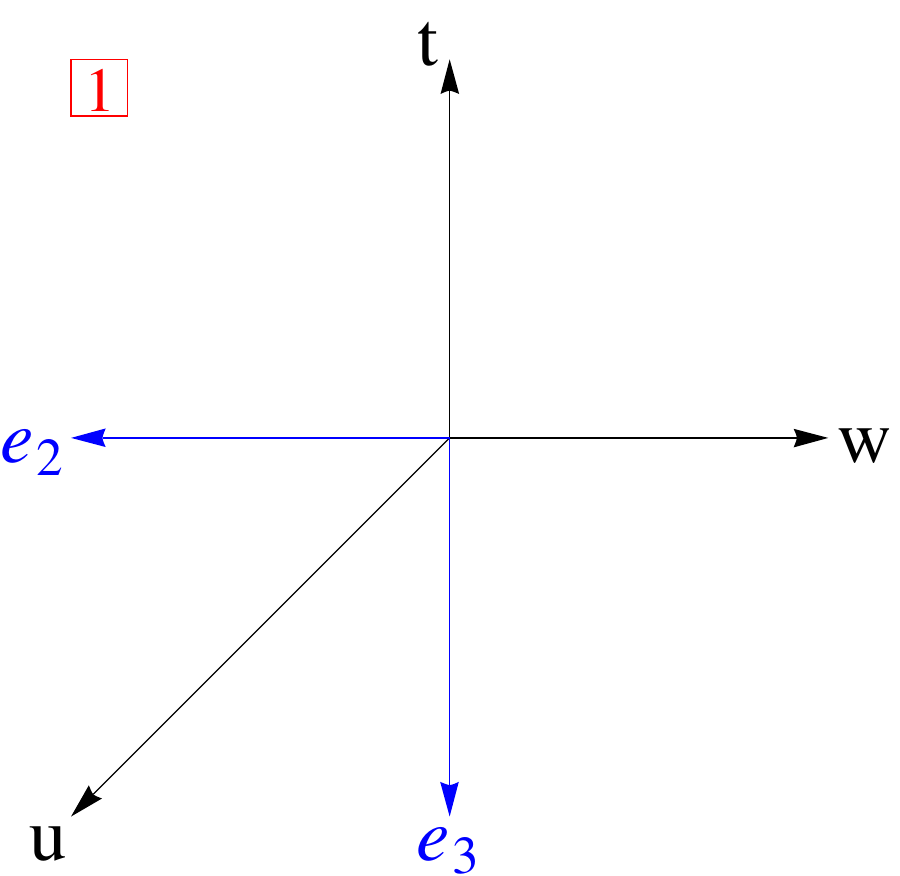}
\includegraphics[scale=0.6]{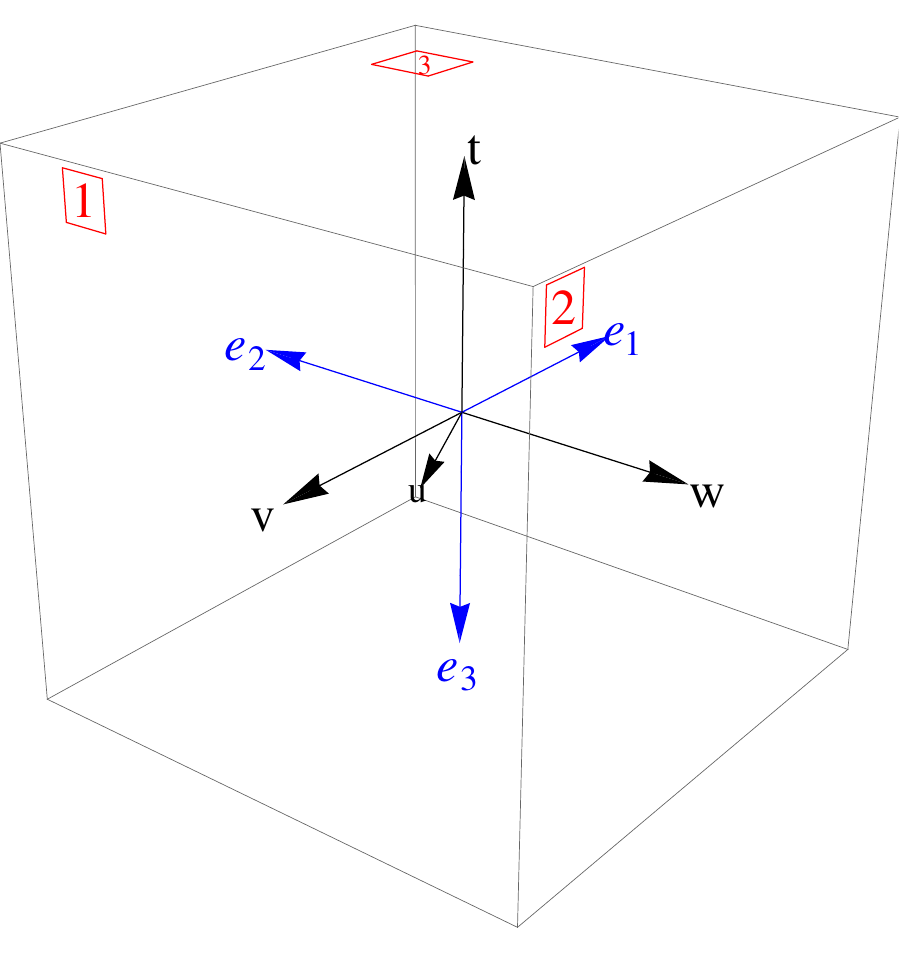}
\includegraphics[scale=0.5]{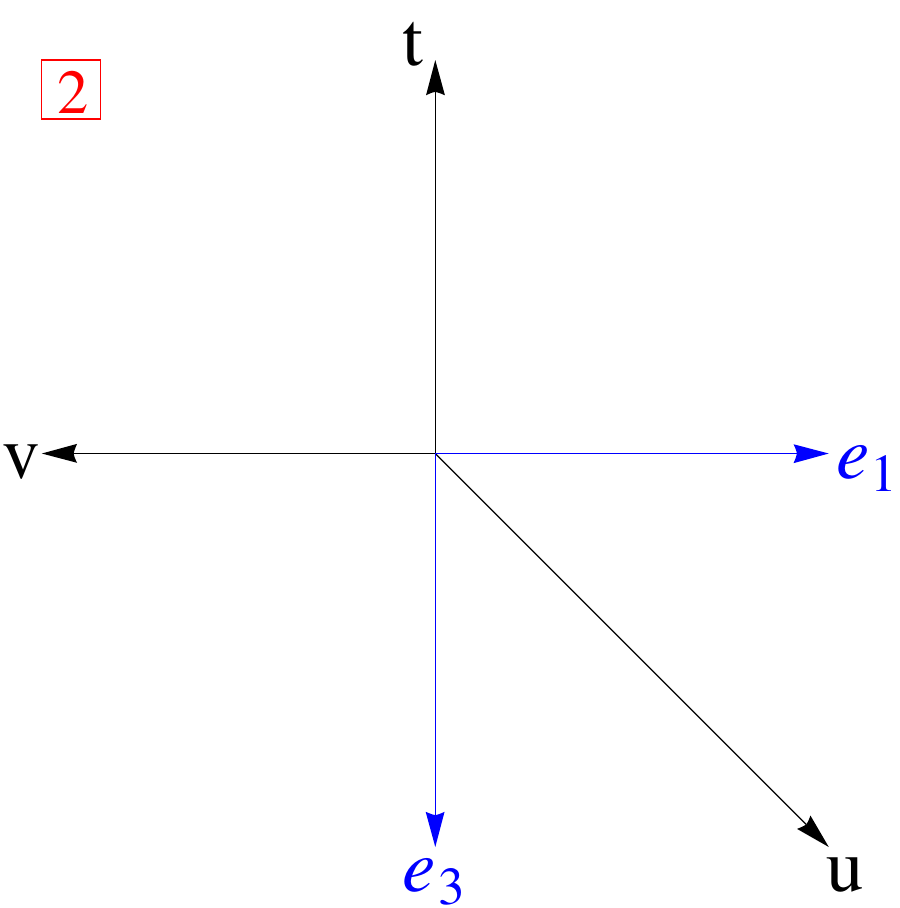}
\caption{Toric fan of $\text{Bl}_3\mathbb{P}^3$ and the 2D projections to the three coordinate planes, each of which yielding the polytope
of $dP_2$.}
\label{fig:BlP3Poly}
\end{figure}

Finally, we show that the presentation of $\mathcal{E}$ as the complete 
intersection \eqref{eq:EonBlP3} can be obtained torically from a 
nef-partition of the $\text{Bl}_3\mathbb{P}^3$. For this purpose we 
only have to realize that the blow-ups \eqref{eq:BlowDownMap} can be 
realized torically by adding the following rays to the polytope of 
$\mathbb{P}^3$ in \eqref{eq:vertsP3},
\beq \label{eq:blowuprays}
\rho_{e_1}=(-1,0,0)\,,\qquad\rho_{e_2}=(0,-1,0)\,,\qquad\rho_{e_3}=(0,0,-1)\,.
\eeq
The rays of the polytope of $\text{Bl}_3\mathbb{P}^3$ are illustrated in 
the center of figure \eqref{fig:BlP3Poly}.

Here the ray $\rho_{e_i}$ precisely corresponds to the exceptional 
divisor $E_i=\{e_i=0\}$. Then we determine the nef-partitions of this
polytope $\nabla_{\text{Bl}_3\mathbb{P}^3}$ of $\text{Bl}_3\mathbb{P}^3$. We find that is admits a single
nef-partition into $\nabla_1$, $\nabla_2$ reading 
\beq \label{eq:nefFiber}
	\nabla_{\text{Bl}_3\mathbb{P}^3}=\langle \nabla_1\cup \nabla_2\rangle\,,\qquad \nabla_1=\langle\rho_1,\rho_4,\rho_{e_1},\rho_{e_2}\rangle\,\quad \nabla_2=\langle\rho_{2},\rho_{3},\rho_{e_3}\rangle\,.
\eeq
It is straightforward to check that the general formula 
\eqref{eq:sectionNEFpart} for the nef-partition at hand reproduces 
precisely the constraints \eqref{eq:EonBlP3}.

\subsection{Connection to the cubic in $dP_2$}
\label{sec:connectionToCubic}

In this section we construct three equivalent maps of the elliptic curve $\mathcal{E}$ given as the intersection \eqref{eq:EonBlP3} in 
$\text{Bl}_3\mathbb{P}^3$ to the Calabi-Yau onefold in $dP_2$. The elliptic curve we obtain will not be the generic elliptic
curve in $dP_2$ found in \cite{Borchmann:2013jwa,Cvetic:2013nia} with rank two Mordell-Weil group, but non-generic with 
a rank three Mordell-Weil group with one \textit{non-toric generator}. The map of the toric generator
of the Mordell-Weil group in $\text{Bl}_3\mathbb{P}^3$ to a  non-toric generator in $dP_2$ will be manifest. 

The presentation of $\mathcal{E}$ as a non-generic hypersurface in $dP_2$ with a non-toric Mordell-Weil group allows us to use the results
of \cite{Cvetic:2013nia} from the analysis of the generic $dP_2$-curve. On the one hand, we can immediately obtain the birational 
map of $\mathcal{E}$ in \eqref{eq:EonBlP3} to the Weierstrass model by first using  the map to $dP_2$  and then by  the map 
from $dP_2$ to the Weierstrass form. We  present this map separately in  section \ref{sec:WSF}. On the other hand, the study of codimension two singularities in section 
\ref{sec:Codim2Sings} will essentially reduce to the analysis of codimension two singularities in fibrations with elliptic fiber in $dP_2$. 
However,  the additional non-toric Mordell-Weil generator as well as the non-generic hypersurface equation in $dP_2$ will give 
rise to a richer structure of codimension two singularities.

\subsubsection{Mapping the Intersection of Two Quadrics in $\mathbb{P}^3$ to the Cubic in $\mathbb{P}^2$}
\label{sec:QuadsToCubic}

As a preparation, we begin with a brief digression on the map of an elliptic curve with a single point $P_0$ given as a 
complete intersection of two quadrics in $\mathbb{P}^3$ to the cubic in $\mathbb{P}^2$, where we closely follow \cite{cassels1991lmsst,connell1996elliptic}. 

Let us assume that there is a rational point $P_0$ on the complete intersection of two  quadrics  with coordinates 
$[x_0:x_1:x_2:x_3]=[0:0:0:1]$ in $\mathbb{P}^3$.\footnote{We choose coordinates $[x_0:x_1:x_2:x_3]$ on $\mathbb{P}^3$ in order to keep our 
discussion here general. We will identify the $x_i$ with the coordinates used in sections \ref{sec:EasQuadricInt} and \eqref{sec:EasIntInBlP3} in
section \ref{sec:QuadsToCubicResolved}.} 
This implies the quadrics must have the form
\beq
\label{eq:QuadToNagell}
A x_3 + B=0\,, \qquad  C x_3 + D=0 \,,
\eeq
where $A$, $C$  are linear and $B$, $D$ are quadratic polynomials in the variables $x_0,\,x_1,\,x_2$. Assuming that $A$, $C$ are generic, we  
obtain a cubic equation in $\mathbb{P}^2$ with coordinates $[x_0:x_1:x_2]$\footnote{We can think of this $\mathbb{P}^2$ as being obtained 
from $\mathbb{P}^3$ via a toric morphism defined by projection along one toric ray. In the case at hand this is the ray corresponding to $x_3=0$.} 
by solving \eqref{eq:QuadToNagell} for $x_3$,
\beq \label{eq:AD-CB}
A D - B C =0\,,
\eeq
Here we have to require that $[x_0:x_1:x_2]\neq [0,0,0]$, because $x_3=-\frac{B}{A}=-\frac{D}{C}$ has to be well-defined. Then, the inverse map
from the cubic in $\mathbb{P}^2$ to the complete intersection \eqref{eq:QuadToNagell} reads
\beq
	[x_0:x_1:x_2]\mapsto [x_0:x_1:x_2:x_3=-\tfrac{B}{A}=-\tfrac{D}{C}]\,.
\eeq
The original point $P_0=[0:0:0:1]$ is mapped to the rational point given by the intersection of the two lines $A=0$, $C=0$.  This can be seen
by noting that $A=C=0$ in \eqref{eq:QuadToNagell} implies also $B=D=0$ which is only solved if $[x_0:x_1:x_2]=[0:0:0]$.

We note that the case when $A$ and $C$ are co-linear, i.e.~$A\sim C$, is special because the curve \eqref{eq:QuadToNagell}
describes no longer a smooth elliptic curve, but a $\mathbb{P}^1$. Indeed, if $A= aC$ for a number $a$ we can rewrite \eqref{eq:QuadToNagell}
as
\beq \label{eq:DegenerationOfQuadrics}
	B-aD=0\,,\qquad Cx_3+D=0\,,
\eeq
where we can solve the second constraint for $x_3$, given $C\neq 0$, so that we are left with the quadratic constraint $B-aD=0$ in 
$\mathbb{P}^2$, 
which is a $\mathbb{P}^1$. This type of degeneration of the complete intersection \eqref{eq:QuadToNagell} will be the prototype for the
degenerations of the elliptic curve \eqref{eq:EonBlP3}, that we find in section \ref{sec:Codim2Sings}.

\subsubsection{Mapping the Intersection in $\text{Bl}_3\mathbb{P}^3$ to the Calabi-Yau Onefold in $dP_2$}
\label{sec:QuadsToCubicResolved}

Next we apply the map of section \ref{sec:QuadsToCubic} to the elliptic curve $\mathcal{E}$ with three rational points. Since  \eqref{eq:1st}
is linear in all three coordinates $v'$, $w'$ and $t'$ we will obtain according to the discussion below \eqref{eq:QuadToNagell}  three canonical 
maps to a cubic in $\mathbb{P}^2$. In fact, these maps lift to maps of the elliptic curve \eqref{eq:EonBlP3} in $\text{Bl}_3\mathbb{P}^3$ to 
elliptic curves presented as Calabi-Yau hypersurfaces in $dP_2$, as we demonstrate in the following.

We construct the map from the complete intersection \eqref{eq:EonBlP3} to the elliptic curve in $dP_2$ explicitly for the point $R$ in 
\eqref{eq:P3coordsPQRS}, i.e.~we identify $P_0\equiv R$ and $[x_0:x_1:x_2:x_3]=[u':v':t':w']$ in the coordinates on $\mathbb{P}^3$ 
before the blow-up for the discussion in section \ref{sec:QuadsToCubic}. Next, we compare \eqref{eq:QuadToNagell} to the complete intersection \eqref{eq:EonBlP3}.  After the blow-up \eqref{eq:BlowDownMap}, the point $R$ is mapped to 
$e_2=0$ as noted earlier in \eqref{eq:resolvedPoints}. This allows us to identify $A$, $C$ in \eqref{eq:QuadToNagell} as those terms  in 
\eqref{eq:EonBlP3}  that do not vanish, respectively, $B$, $D$ as the terms that vanish for $e_2=0$. Thus we effectively rewrite
\eqref{eq:EonBlP3} in the form \eqref{eq:QuadToNagell} with $x_3\equiv w$ after the blow-up, since $w=1$ follows from \eqref{eq:SRBlP3} for 
$e_2=0$,  and obtain
\bea \label{eq:ACQuadNagell}
\phantom{.}A=   s_7 e_1e_3u + s_8 e_3v -s_{10}e_1 t\,, \quad C =  s_{17} e_1e_3u + s_{18} e_3v-s_{20} e_1t\,, \hspace{1.8cm}\\
\!B=e_2(s_2e_1 e_3 u^2+s_5e_1 ut+s_{6} e_3 u v  -s_9  vt )\,,\quad\!\! D=e_2(s_{12}e_1  e_3 u ^2 + s_{15}e_1   ut + s_{16} e_3 u v  -s_{19}  vt)\,.\!\!\!\!\!\!\!\!\!\!\!\!\!\nn
\eea
In particular, this identification implies that $R=\{e_2=0\}$ is mapped to $A=C=0$ on $dP_2$ as required.
Then, we solve both equations for $w$ and obtain the hypersurface equation of the form
\beq
\label{eq:cubicP2}
u(\tilde{s}_1 u^2 e_1^2e_3^2 +\tilde{s}_2 u v e_1 e_3^2+\tilde{s}_3  v^2e_3^2 + \tilde{s}_5 u t e_1^2 e_3+ \tilde{s}_6  v t e_1 e_3+ \tilde{s}_8  t^2 e_1^2 ) + \tilde{s}_7 v^2 t e_3+ \tilde{s}_9 v t^2 e_1= 0 \,,
\eeq
where we have set $e_2=1$ using one $\mathbb{C}^*$-action on $\text{Bl}_3\mathbb{P}^3$ as $B,\, D\sim e_2$ and $e_2=0$ implies 
$w=-\frac{B}{A}=-\frac{D}{C}=0$ which is inconsistent with the SR-ideal 
\eqref{eq:SRBlP3} . The coefficients $\tilde{s}_i$ in \eqref{eq:cubicP2} read
\beq \label{eq:sP3ToP2I}
\text{
\begin{tabular}{|c|c|}
\hline
 &   coefficients in $dP_2$-curve projected along $[w:e_2]$\\ \hline
$\tilde{s}_1$ & $-s_{17} s_{2} + s_{12} s_{7}$  \\
$\tilde{s}_2$ & $-s_{18} s_2 - s_{17} s_6 + s_{16} s_7 + s_{12} s_{8}$ \\
$\tilde{s}_3$ & $-|M_1^Q|=s_{16} s_{8}-s_{18} s_6 $ \\
$\tilde{s}_5$ & $-s_{10} s_{12} + s_2 s_{20} - s_{17} s_5 + s_{15} s_7$  \\ 
$\tilde{s}_6$ & $-s_{10} s_{16} - s_{18} s_5 + s_{20} s_6 - s_{19} s_7 + s_{15} s_8 + s_{17} s_9$\\ 
$\tilde{s}_7$ & $|M_3^Q|=s_{18} s_{9}-s_{19} s_{8}  $\\ 
$\tilde{s}_8$ & $-|M_2^P|=-s_{10} s_{15} + s_{20} s_{5}$\\ 
$\tilde{s}_9$ & $-|M_3^P|=s_{10} s_{19} - s_{20} s_{9}$\\ \hline
\end{tabular}
}
\eeq
Here we have used the minors introduced in \eqref{eq:MSDets} and in \eqref{eq:PQRSMinors}, \eqref{eq:MatsPQR}.

We note that the ambient space of  \eqref{eq:cubicP2} is  $dP_2$ with homogeneous coordinates $[u:v:w:t:e_1:e_3]$. The relevant $dP_2$ is 
obtained from $\text{Bl}_3\mathbb{P}^3$ by a toric morphism that is defined by projecting the polytope of  $\text{Bl}_3\mathbb{P}^3$ generated 
by \eqref{eq:vertsP3}, \eqref{eq:blowuprays} onto the plane that is perpendicular to the line through the rays $\rho_3$ and  $\rho_{e_2}$. The rays of the fan are shown in the figure on the right of \ref{fig:BlP3Poly} that is obtained by the projection of the rays on the face number two of the cube.  This can also
be seen from the unbroken $\mathbb{C}^*$-actions in \eqref{eq:divclassBlP3} and the SR-ideal  \eqref{eq:SRBlP3} for $e_2=1$ and $w=0$, or 
$e_2=0$ and $w=1$.
Then, the cubic \eqref{eq:cubicP2} is a section precisely of the anti-canonical bundle of this $dP_2$ surface. 

The general elliptic curve in $dP_2$ was studied in \cite{Cvetic:2013nia,Borchmann:2013jwa} and shown to have a rank two Mordell-Weil group. 
However, the elliptic curve \eqref{eq:cubicP2} has by construction a rank three Mordell-Weil group. Indeed, we see that the coefficients $\tilde{s}_i$ 
are non-generic and precisely allow for a fourth rational point. This fourth point, however, does not descend from a divisor of the ambient space
$dP_2$ and is not toric. In fact, the mapping of the four rational points \eqref{eq:resolvedPoints} in the coordinates on $dP_2$  reads 
\bea \label{eq:mappedPoints}
P=[-|M^P_3|:|M^P_2|:-|M^P_1| :1:1:1:0] \phantom{} &\mapsto &\,  [|M^P_3|:-|M^P_2| :1:1:0] \,,    \\
Q=[-|M^Q_3|:1:|M^Q_2|:-|M^Q_1|:0:1:1]   &\mapsto  & \, [-|M^Q_3|:1:-|M^Q_1|:0:1] \,,\nn   \\
R=[|M^R_3|:-|M^R_2|:1 :|M^R_1|:1:0:1]\,\,\, \,\, &\mapsto  &\, [|M^R_3|:-|M^R_2| :|M^R_1|:1:1] \,,    \nn \\
S=[ 0:1:1 :1:-|M^Q_3|: |M^R_3|:-|M^P_3|] &\mapsto &\,   [0:1 :1:-|M^Q_3|: -|M^P_3|] \,.   \nn 
\eea
We see, that the points $P$, $Q$ and  $S$ are mapped to the three toric points on the elliptic curve in $dP_2$ studied in \cite{Cvetic:2013nia},
whereas the points $R$ is mapped to a non-toric point.

The map from the complete intersection in $\text{Bl}_3\mathbb{P}^3$ to the elliptic curve \eqref{eq:cubicP2} in $dP_2$ implies that the results 
from the analysis of \cite{Cvetic:2013nia}, where the generic elliptic curve in $dP_2$ was considered, immediately apply.
More precisely, renaming the coordinates $[u:v:t:e_1:e_3]$ in \eqref{eq:cubicP2} as $[u:v:w:e_1:e_2]$ we readily recover equation (3.4) of 
\cite{Cvetic:2013nia}. Furthermore, the points $P$, $Q$ and $S$ in \eqref{eq:mappedPoints} immediately map to the origin  and the two rational 
points of the rank two elliptic curve in $dP_2$, that we denote in the following as $\tilde{P}$, $\tilde{Q}$ and $\tilde{R}$. In the notation of 
\cite{Cvetic:2013nia} we thus rewrite \eqref{eq:mappedPoints} using \eqref{eq:sP3ToP2I} as
\bea \label{eq:PQStoTildePQR}
	&P\,\,\,\mapsto\,\,\, \tilde{P}:=[-\tilde{s}_9:\tilde{s}_8:1:1:0]\,,\quad Q\,\,\,\mapsto\,\,\, \tilde{Q}:=[-\tilde{s}_7:1:\tilde{s}_3:0:1]\,,&\nn\\
	& S\,\,\,\mapsto\,\,\, \tilde{R}:=[0:1:1:-\tilde{s}_7:\tilde{s}_9]\,.&
\eea
We emphasize that the origin $P$ in the complete intersection in \eqref{eq:EonBlP3} is mapped to the origin $\tilde{P}$, which implies that the
Weierstrass form of the curve in $dP_2$ will agree with the Weierstrass form of the curve \eqref{eq:EonBlP3}, cf.~section \ref{sec:WSF}.

As we mentioned before, the point $R$ is mapped to a non-toric point in $dP_2$. This complicates the determination of the Weierstrass
coordinates for $R$, for example.  Fortunately, there are two other maps of the elliptic curve 
\eqref{eq:EonBlP3} to a curve in $dP_2$ in which the point $R$ is mapped to a toric point and another point, either  $Q$ or $P$, are realized
non-torically. Thus, we construct in the following a second map to an elliptic curve
in $dP_2$, where $R$ is toric. Since the logic is completely analogous to the previous construction, we will be as brief as possible.

We choose $P_0\equiv Q$ for the map to $dP_2$. We recall from \eqref{eq:resolvedPoints} that $Q$ is realized as $e_1=0$ on the elliptic curve in 
$\text{Bl}_3\mathbb{P}^3$. Thus, we write \eqref{eq:EonBlP3} as
\beq \label{eq:QuadToNagell2}
A v+ B=0 \,, \qquad Cv + D=0 \,,
\eeq
where, as before, $A$ and $C$ are obtained by setting $e_1=0$ and $B$, $D$ are  the terms proportional to $e_1$,
\bea
	&A= -s_9e_2 t + s_6 e_2e_3u + s_8 e_3w \,, \quad C = -s_{19}e_2 t + s_{16}e_2e_3 u + s_{18} e_3w \,,&\\
	&\!\!B=e_1(s_2 e_2 e_3 u^2+s_5 e_2  ut +s_7 e_3 u w  -s_{10}  wt) \,,\! \!\quad D=e_1(s_{12} e_2 e_3 u^2+s_{15} e_2  ut +s_{17} e_3 u w  -s_{20}  wt) \,.&\nn
\eea 
Thus, we obtain an elliptic curve in $dP_2$  with homogeneous coordinates $[u:w:t:e_2:e_3]$ by solving \eqref{eq:QuadToNagell2} for $v$ and
by setting $e_1=1$ as required by the SR-ideal \eqref{eq:SRBlP3}. The hypersurface constraint \eqref{eq:AD-CB} takes the form
\beq
\label{eq:cubicP2_v}
u(\hat{s}_1 u^2 e_2^2e_3^2 +\hat{s}_2 u w e_2 e_3^2+\hat{s}_3  w^2e_3^2 + \hat{s}_5 u t e_2^2 e_3+ \hat{s}_6  w t e_2 e_3+ \hat{s}_8  t^2 e_2^2 ) + \hat{s}_7 w^2 t e_3+ \hat{s}_9 w t^2 e_2= 0 \,, 
\eeq
with coefficients $\hat{s}_i$ defined as
\beq \label{eq:sP3ToP2II}
\text{
\begin{tabular}{|c|c|}
\hline
 &  coefficients in $dP_2$-curve projected along $[v:e_1]$  \\ \hline
$\hat{s}_1$ & $-s_{16} s_{2} + s_{12} s_{6}$  \\
$\hat{s}_2$ & $-s_{18} s_2 + s_{17} s_6 - s_{16} s_7 + s_{12} s_{8}$ \\
$\hat{s}_3$ & $-|M_1^R|=-s_{18} s_7 + s_{17} s_{8}$ \\
$\hat{s}_5$ & $s_{19} s_{2} - s_{16} s_{5} + s_{15} s_6 - s_{12} s_9$  \\ 
$\hat{s}_6$ & $s_{10} s_{16} - s_{18} s_5 - s_{20} s_6 + s_{19} s_7 + s_{15} s_8 - s_{17} s_9$\\ 
$\hat{s}_7$ & $|M_3^R|=s_{10} s_{18} - s_{20} s_{8}$\\ 
$\hat{s}_8$ & $-|M_1^P|=s_{19} s_{5} - s_{15} s_{9}$\\ 
$\hat{s}_9$ & $|M_3^P|=-\tilde{s}_9=-s_{10} s_{19} + s_{20} s_{9}$\\ \hline
\end{tabular}
}
\eeq
where we have used \eqref{eq:sP3ToP2I}. Analogously to the previous map, the ambient space of the hypersurface \eqref{eq:cubicP2_v}
is the $dP_2$ with homogeneous coordinates $[u:w:t:e_2:e_3]$ that is obtained from $\text{Bl}_3\mathbb{P}^3$ by the toric morphism
induced by projecting along the line through the rays $\rho_2$ and $\rho_{e_1}$. The rays of the fan are shown in the left figure of \ref{fig:BlP3Poly} that corresponds to the projection of the rays on the face number one.
Then, the three rational points on $\mathcal{E}$ and the origin get mapped, in the coordinates $[u:w:t:e_2:e_3]$ of $dP_2$, to
\bea \label{eq:PQRSondP2_v}
P=[-|M^P_3|:|M^P_2|:-|M^P_1| :1:1:1:0] \phantom{} &\mapsto &\,  [-|M^P_3|:-|M^P_1| :1:1:0] \,,    \\
Q=[-|M^Q_3|:1:|M^Q_2|:-|M^Q_1|:0:1:1]   &\mapsto  & \, [-|M^Q_3|:|M^Q_2|:-|M^Q_1|:1:1] \,,\nn   \\
R=[|M^R_3|:-|M^R_2|:1 :|M^R_1|:1:0:1]\,\,\, \,\, &\mapsto  &\, [|M^R_3|:1 :|M^R_1|:0:1] \,,    \nn \\
S=[ 0:1:1 :1:-|M^Q_3|: |M^R_3|:-|M^P_3|] &\mapsto &\,   [0:1 :1:|M^R_3|:-|M^P_3|] \,.   \nn 
\eea

As before, it is convenient to make contact to the notation  of \cite{Cvetic:2013nia}. After  the renaming $[u:w:t:e_2:e_3]\rightarrow 
[u:v:w:e_1:e_2]$ we obtain the hypersurface constraint \eqref{eq:cubicP2_v} takes the standard form of eq. (3.4) in \cite{Cvetic:2013nia}. In
addition, we see that the points $P$, $R$ and $S$ get mapped to the toric points on $dP_2$, whereas $Q$ maps to a non-toric point. Denoting
the origin of the $dP_2$-curve by $\hat{P}$ and the two rational points by $\hat{Q}$, $\hat{R}$ in order to avoid confusion, we then write 
\eqref{eq:PQRSondP2_v} as
\bea \label{eq:PRStocubic}
	&P\,\,\,\mapsto\,\,\, \hat{P}:=[-\hat{s}_9:\hat{s}_8:1:1:0]\,,\quad R\,\,\,\mapsto\,\,\, \hat{Q}=[-\hat{s}_7:1:\hat{s}_3:0:1]\,,&\nn\\
	& S\,\,\,\mapsto\,\,\, \tilde{R}=[0:1:1:\hat{s}_7:-\hat{s}_9]\,.&
\eea

We note that there is a third map from \eqref{eq:EonBlP3} to $dP_2$ by solving for the variable $t$, respectively, $e_3$ (its fan would correspond to the upper figure in figure \ref{fig:BlP3Poly} that shows the projection of the rays in the face number three). Although this map
is formally completely analogous to the above the maps, it is not very illuminating for our purposes since the chosen zero point $P$
on $\mathcal{E}$ maps to a non-toric point in $dP_2$. In particular, the Weierstrass model with respect to $P$ can not be obtained
from this elliptic curve in $dP_2$ by simply applying the results of \cite{Cvetic:2013nia}, where $P$ by assumption has to be a toric point.

\subsection{Weierstrass Form with Three Rational Points}
\label{sec:WSF}

Finally, we are prepared to obtain the Weierstrass model for the elliptic curve $\mathcal{E}$ in \eqref{eq:EonBlP3} with respect to the chosen
origin $P$ along with the coordinates in Weierstrass form for the three rational points $Q$, $R$ and $S$. We present three maps to a Weierstrass 
model in this work, each of which yielding an identical Weierstrass form, i.e.~identical $f$, $g$ in $y^2=x^3+f xz^4+gz^6$.  The details of the relevant computations as well as the explicit results can be found in appendix \ref{app:WSFdirectly}.

The simplest two ways to obtain this Weierstrass from is by first exploiting the two presentations of the elliptic curve $\mathcal{E}$ as the 
hypersurfaces \eqref{eq:cubicP2} and \eqref{eq:cubicP2_v} in  $dP_2$ constructed in section \ref{sec:QuadsToCubicResolved} and by then using 
the birational map of \cite{Cvetic:2013nia} of the general elliptic curve in $dP_2$ to the Weierstrass form in $\mathbb{P}^{2}(1,2,3)$. 
In summary, we find the following schematic coordinates for the coordinates in Weierstrass form of the  rational points $Q$, $R$ and $S$
\beq \label{eq:QRSgeneralForm}
	Q=[g_2^Q:g_3^Q:1]\,,\quad R=[g_2^R:g_3^R:1]\,,\qquad S=[g_2^S:g_3^S:(s_{10}s_{19}-s_{9}s_{20})]
\eeq
with the explicit expressions for $g_2^{Q,R,S}$ and $g_3^{Q,R,S}$ given in (\ref{eq:xyQ}-\ref{eq:xyS}) in appendix \ref{app:WSFdirectly}.  The
explicit form for $f$ and $g$, along with the discriminant follow from the formulas in \cite{Cvetic:2013nia} in combination with 
\eqref{eq:sP3ToP2I}, respectively, \eqref{eq:sP3ToP2II}. 
In fact, we obtain \eqref{eq:QRSgeneralForm} for $Q$ and $S$ by using the presentation \eqref{eq:cubicP2} 
along with the maps \eqref{eq:PQStoTildePQR} of the rational points $Q$ and $S$ onto the two toric points in the $dP_2$-elliptic curve, denoted 
by $\tilde{Q}$ and $\tilde{R}$ in this context. 
Then, we apply  Eqs.~(3.11) and (3.12)  of \cite{Cvetic:2013nia}  for the coordinates in 
Weierstrass form of the two toric rational points on the elliptic curve in $dP_2$. For concreteness, for the curve \eqref{eq:cubicP2} the coordinates
in Weierstrass form of the two points read
\beq
 [g_2^Q:g_3^Q:z_Q]=   [\tfrac{1}{12} (\tilde{s}_6^2 - 4 \tilde{s}_5 \tilde{s}_7 + 8 \tilde{s}_3 \tilde{s}_8 - 4 \tilde{s}_2 \tilde{s}_9),\tfrac{1}{2} (\tilde{s}_3 \tilde{s}_6 \tilde{s}_8 - \tilde{s}_2 \tilde{s}_7 \tilde{s}_8 - \tilde{s}_3 \tilde{s}_5 \tilde{s}_9 + \tilde{s}_1 \tilde{s}_7 \tilde{s}_9):1]\,
 \label{eq:MWGQ}
\eeq
for the point $\tilde{Q}=[-\tilde{s}_7:1:\tilde{s}_3:0:1]$ and 
\bea 
  g_2^S &=&  \tfrac{1}{12} (12 \tilde{s}_7^2 \tilde{s}_8^2 + \tilde{s}_9^2 (\tilde{s}_6^2 + 8 \tilde{s}_3 \tilde{s}_8 - 4 \tilde{s}_2 \tilde{s}_9) + 
 4 \tilde{s}_7 \tilde{s}_9 (-3 \tilde{s}_6 \tilde{s}_8 + 2 \tilde{s}_5 \tilde{s}_9))\,, \nn \\ 
     g_{3}^S&=& \tfrac{1}{2} (2 \tilde{s}_7^3 \tilde{s}_8^3 + \tilde{s}_3 \tilde{s}_9^3 (-\tilde{s}_6 \tilde{s}_8 + \tilde{s}_5 \tilde{s}_9) + 
 \tilde{s}_7^2 \tilde{s}_8 \tilde{s}_9 (-3 \tilde{s}_6 \tilde{s}_8 + 2 \tilde{s}_5 \tilde{s}_9)  \nn \\ 
 & &+\tilde{s}_7 \tilde{s}_9^2 (\tilde{s}_6^2 \tilde{s}_8 + 2 \tilde{s}_3 \tilde{s}_8^2 - \tilde{s}_5 \tilde{s}_6 \tilde{s}_9 - \tilde{s}_2 \tilde{s}_8 \tilde{s}_9 + \tilde{s}_1 \tilde{s}_9^2)\,, \nn \\ 
    z_{S}&=& \tilde{s}_9  \,
    \label{eq:MWGR}
\eea
for the point $\tilde{R}=[0:1:1-\tilde{s}_7:\tilde{s}_9]$, where we apply \eqref{eq:sP3ToP2I}.
The explicit result in terms of the coefficients $s_i$ for both $Q$, $S$ can be found in \eqref{eq:xyQ}, respectively, \eqref{eq:xyS}. 

In order to 
obtain the Weierstrass coordinates for the point $R$ in  \eqref{eq:QRSgeneralForm} we invoke the map 
$R\mapsto \hat{Q}$ in \eqref{eq:PRStocubic} for the elliptic curve \eqref{eq:cubicP2_v} in $dP_2$. Here, the coordinates of 
$R\mapsto \hat{Q}$ are again given by \eqref{eq:MWGQ} after replacing $\tilde{s}_i\rightarrow \hat{s}_i$. The explicit 
form for these coordinates in terms of the $s_i$ is obtained using \eqref{eq:sP3ToP2II} and can be found in \eqref{eq:xyR}. 
We emphasize that the coordinates in Weierstrass form  for $S$
can also be obtained from the map $S\mapsto \hat{R}$ in \eqref{eq:PRStocubic} in combination with \eqref{eq:sP3ToP2II}. They precisely 
agree with those in \eqref{eq:xyS} deduced from the map $S\mapsto \tilde{R}$ and \eqref{eq:sP3ToP2I}.

Alternatively,
one can directly construct the birational map from \eqref{eq:EonBlP3} to the Weierstrass form by extension of  the techniques of 
\cite{Morrison:2012ei,Cvetic:2013nia}, where $x$ and $y$  in $\mathbb{P}^{2}(1,2,3)$ are constructed as sections of appropriate line bundles that
vanish with appropriate degrees at $Q$, $R$ and $S$. However, the corresponding calculations are lengthy and the resulting Weierstrass model 
is identical to the one obtained from $dP_2$. Thus, we have opted to relegate this analysis to appendix \ref{app:WSFdirectly}.

\section{Elliptic Fibrations with Three Rational Sections}
\label{sec:EllipticFibrations}

In this section we construct resolved elliptically fibered Calabi-Yau 
manifolds $\mathcal{E} \rightarrow\hat{X}\stackrel{\pi}{\rightarrow} B$ over a base $B$ with a rank three Mordell-Weil group.
The map $\pi$ denotes the projection to the base $B$ and the general elliptic fiber $\mathcal{E}=\pi^{-1}(pt)$ over a generic point $pt$ in $B$ is 
the elliptic curve with rank three Mordell-Weil group of section \ref{sec:3waysToRk3}. An elliptic Calabi-Yau manifold $\hat{X}$ with all 
singularities at higher codimension resolved is obtained by fibering $\mathcal{E}$ in the presentation \eqref{eq:EonBlP3}. In addition, in this 
representation for $\mathcal{E}$ the generators of the Mordell-Weil group are given by the restriction to $\hat{X}$ of the toric divisors of the 
ambient space $\text{Bl}_3\mathbb{P}^3$ of the fiber, i.e.~the Mordell-Weil group of the generic $\hat{X}$ is toric.

We begin  in section \ref{sec:ConstructionEllipticFibrations} with the construction of Calabi-Yau elliptic fibrations $\hat{X}$ with rank three
Mordell-Weil group over a general base $B$ with the elliptic curve \eqref{eq:EonBlP3}  as the general elliptic fiber. We see that all these
fibrations are classified by three divisors in the base $B$.  Then in section \ref{sec:BasicGeometry} we compute the universal 
intersections on $\hat{X}$, that hold generically and are valid for any base $B$. Finally, in section \ref{sec:allVacua} we classify all generic
Calabi-Yau manifolds $\hat{X}$ with elliptic fiber $\mathcal{E}$ in $\text{Bl}_3\mathbb{P}^3$ over any base $B$. Each such F-theory
vacua $\hat{X}$ is labeled by one point in a particular polytope, that we determine. 

The techniques and results in the following analysis are a direct extension to the ones used in  
\cite{Cvetic:2013nia,Cvetic:2013uta,Cvetic:2013jta} for the case of a rank two Mordell-Weil group.

\subsection{Constructing Calabi-Yau Elliptic Fibrations}
\label{sec:ConstructionEllipticFibrations}

Let us begin with the explicit construction of the Calabi-Yau manifold $\hat{X}$. Abstractly, a general elliptic fibration of the given elliptic curve 
$\mathcal{E}$ over a base $B$ is given by defining the complete intersection \eqref{eq:EonBlP3}  over the function 
field of $B$. In other words, we lift  all coefficients $s_i$ as well as the coordinates in \eqref{eq:EonBlP3} to sections of appropriate line bundles 
over $B$. 

To each of the homogeneous coordinates on $\text{Bl}_3\mathbb{P}^3$ we assign a different line bundle on the base $B$. However,
we can use the $(\mathbb{C}^*)^4$-action in \eqref{eq:divclassBlP3} to assign without loss of generality  the following non-trivial line bundles
\beq \label{eq:LBassignment}
	u\in \mathcal{O}_B(D_u)\,,\qquad v\in \mathcal{O}_B(D_v)\,,\qquad w\in \mathcal{O}_B(D_w)\,,
\eeq
with all other coordinates $[t:e_1:e_2:e_3]$ transforming in the trivial bundle on $B$.
Here $K_B$ denotes the canonical bundle on $B$, $[K_B]$ the
associated divisor and $D_u$, $D_v$ and $D_w$ are three, at the moment, arbitrary divisors on $B$. They will be fixed 
later in this section by the Calabi-Yau condition on the elliptic fibration.
The assignment \eqref{eq:LBassignment} can be described globally by constructing the fiber bundle
\beq \label{eq:BlP3fibration}
	\xymatrix{
	\text{Bl}_3 \mathbb{P}^3 \ar[r] & 	\BPF  \ar[d]\\
	& B\,	}
\eeq
The total space of this fibration is the ambient space of the complete intersection \eqref{eq:EonBlP3}, that defines the elliptic fibration 
of $\mathcal{E}$ over $B$.

Next, we require the complete intersection \eqref{eq:EonBlP3} to define a Calabi-Yau manifold in the ambient space \eqref{eq:BlP3fibration}.
To this end, we first calculate the anti-canonical bundle of $\BPF$
via adjunction. We obtain
\beq
K^{-1}_{\BPFs}=4H-2E_1-2E_2-2E_3+[K_B^{-1}]+D_u+D_v+D_w\,,
\eeq
where we suppressed the dependence on the vertical divisors $D_u$, $D_v$ and $D_w$ for brevity of our notation and
$H$ as well as the $E_i$ are the classes introduced in \eqref{eq:divclassBlP3}.
For the complete intersection \eqref{eq:EonBlP3} to define a Calabi-Yau manifold $\hat{X}$ in \eqref{eq:BlP3fibration} we infer again from 
adjunction that the sum of the classes of the two constraints $p_1$, $p_2$ has to be agree with  $[K^{-1}_{\BPFs}]$. Thus, 
the Calabi-Yau condition
reads
\beq \label{eq:CYcondition}
	[p_1]+[p_2]\stackrel{!}{=}4H-2E_1-2E_2-2E_3+[K_B^{-1}]+D_u+D_v+D_w\,.
\eeq
We see from \eqref{eq:divclassBlP3} that both constraints in \eqref{eq:EonBlP3} are automatically in the divisor class $2H-E_1-E_2-E_3$ w.r.t.~the 
classes on the fiber $\text{Bl}_3\mathbb{P}^3$. Thus, \eqref{eq:CYcondition} effectively reduces to a condition on the class of \eqref{eq:EonBlP3} 
in the homology of the base $B$. Denoting the part of the homology classes of the $[p_i]$ in the base $B$ by $[p_1]^b$ and $[p_2]^b+D_v+D_w$, 
we obtain 
\beq \label{eq:CYconditionBase}
	[p_1]^b+[p_2]^b\stackrel{!}{=}[K_B^{-1}]+D_u\,.
\eeq
Here we shifted the class $[p_2]^b\rightarrow D_v+D_w+[p_2^{b}]$ for reasons that will become clear in section \ref{sec:allVacua}.

Using this information we fix the  line bundles on $B$ in which the coefficients $s_i$ take values. We infer from \eqref{eq:EonBlP3}, 
\eqref{eq:LBassignment} and the Calabi-Yau condition \eqref{eq:CYconditionBase} the following assignments of line bundles,
\beq \label{eq:sectionsFibration}
\text{
\begin{tabular}{c|c}
\text{section} & \text{line-bundle}\\
\hline
	$s_2$&$\mathcal{O}([K_B^{-1}] - D_u - [p_2]^b)$\rule{0pt}{13pt} \\
	$s_5$&$\mathcal{O}([K_B^{-1}]-[p_2]^b)$\rule{0pt}{12pt} \\
	$s_6$&$\mathcal{O}([K_B^{-1}]-[p_2]^b-D_v)$\rule{0pt}{12pt} \\
	$s_7$&$\mathcal{O}([K_B^{-1}]-[p_2]^b-D_w)$\rule{0pt}{12pt} \\
	$s_8$&$\mathcal{O}([K_B^{-1}]-[p_2]^b+D_u-D_v-D_w)$\rule{0pt}{12pt} \\
	$s_9$&$\mathcal{O}([K_B^{-1}]-[p_2]^b+D_u-D_v)$\rule{0pt}{12pt} \\
	$s_{10}$&$\mathcal{O}([K_B^{-1}]-[p_2]^b+D_u-D_w)$\rule{0pt}{12pt} \\
\end{tabular}
}\qquad \text{
\begin{tabular}{c|c}
\text{section} & \text{line-bundle}\\
\hline
	$s_{12}$&$\mathcal{O}(-2 D_u + D_v + D_w + [p_2]^b)$\rule{0pt}{13pt} \\
	$s_{15}$&$\mathcal{O}(-D_u + D_v + D_w + [p_2]^b)$\rule{0pt}{12pt} \\
	$s_{16}$&$\mathcal{O}(-D_u + D_w + [p_2]^b)$\rule{0pt}{12pt} \\
	$s_{17}$&$\mathcal{O}(-D_u + D_v + [p_2]^b)$\rule{0pt}{12pt} \\
	$s_{18}$&$\mathcal{O}([p_2]^b)$\rule{0pt}{12pt} \\
	$s_{19}$&$\mathcal{O}(D_w +[p_2]^b)$\rule{0pt}{12pt} \\
	$s_{20}$&$\mathcal{O}(D_v+[p_2]^b)$\rule{0pt}{12pt} \\
\end{tabular}
}
\eeq
We also summarize the complete line bundles of the homogeneous coordinates on $\text{Bl}_3\mathbb{P}^3$ by combining 
the classes in \eqref{eq:divclassBlP3}  and \eqref{eq:LBassignment},
\beq \label{eq:FullClassesuvwt}
\text{
\begin{tabular}{c|c}
\text{section} & \text{bundle}\\
\hline
	$u$&$\mathcal{O}(H-E_1-E_2-E_3+D_u)$\rule{0pt}{13pt} \\
	$v$&$\mathcal{O}(H-E_2-E_3+D_v)$\rule{0pt}{12pt} \\
	$w$&$\mathcal{O}(H-E_1-E_3+D_w)$\rule{0pt}{12pt} \\
	$t$&$\mathcal{O}(H-E_1-E_2)$\rule{0pt}{12pt} \\
	$e_1$&$\mathcal{O}(E_1)$\rule{0pt}{12pt} \\
	$e_2$&$\mathcal{O}(E_2)$\rule{0pt}{12pt} \\
	$e_3$&$\mathcal{O}(E_3)$\rule{0pt}{12pt} \vspace{1.0cm}\\
\end{tabular}
}
\eeq

For later reference,  we point out that the divisors associated to the 
vanishing of the coefficients $\tilde{s}_7$, $\hat{s}_7$ and $\tilde{s}_9=-\hat{s}_9$, denoted as   $\tilde{\cS}_7$,  $\hat{\cS}_7$ respectively 
$\cS_9$, in the two presentations \eqref{eq:cubicP2} and \eqref{eq:cubicP2_v}  in $dP_2$ of the  elliptic curves $\mathcal{E}$ are given by 
\bea \label{eq:S7S9divisors}
&\tilde{\cS}_7:= [-s_{19}s_8+s_{18} s_9] = [K_B^{-1}]+D_u-D_v\,, \quad\!
\hat{\cS}_7: = \left[s_{10}s_{18}-s_{20} s_8\right] =[K_B^{-1}]+D_u-D_w\,,&\!\!\!\!\!\!\nn\\
&\cS_9:=\left[\tilde{s}_9\right]=\left[\hat{s}_9\right] = \left[-s_{10}s_{19}+s_{20} s_9\right] =D_u+[K_B^{-1}] \,.&\!\!\!\!\!\!
\eea
Here we have used the definitions in \eqref{eq:sP3ToP2I}, respectively,
\eqref{eq:sP3ToP2II} together with \eqref{eq:sectionsFibration}  and denoted the divisor classes of a section $s_i$ by $[\cdot]$. 

It is important to notice that the line bundles of the $s_i$ admit an 
additional degree of freedom due to the choice of the class $[p_2]^b$, 
the divisor class of the second constraint $p_2$ in the homology of $B$. This is due to the fact that the Calabi-Yau condition \eqref{eq:CYconditionBase} is 
a partition problem, that only fixes the sum  of the classes $[p_1]^b$, $[p_2]^b$ but leaves the individual classes undetermined. 
For example, in complete intersections in a toric ambient space \eqref{eq:BlP3fibration} the freedom of the class $[p_2]^b$ is fixed by finding
all nef-partitions of the toric polytope associated to 
\eqref{eq:BlP3fibration} that are consistent with the nef-partition 
\eqref{eq:nefFiber} of the $\text{Bl}_3\mathbb{P}^3$-fiber. We 
discuss the freedom in $[p_2]^b$ further in section \ref{sec:allVacua}.

\subsection{Basic Geometry of Calabi-Yau Manifolds with $\text{Bl}_3 \mathbb{P} ^3$-elliptic
Fiber}
\label{sec:BasicGeometry}

Let us next discuss the basic topological properties of the Calabi-Yau manifold $\hat{X}$. 

We begin by constructing a basis $D_A$ of the group of divisors 
$H^{(1,1)}(\hat{X})$ on $\hat{X}$ that is convenient for the study
of F-theory on $\hat{X}$.
A basis of divisors on the generic complete intersection $\hat{X}$ is induced from
the basis of divisors of the ambient space $\text{Bl}_3 \mathbb{P}^3(\tilde{\cS}_7,\hat{\cS}_7,\cS_9)$ by restriction
to $\hat{X}$.
There are the vertical divisors $D_\alpha$ that are obtained by pulling back divisors $D_\alpha^b$ on the base $B$ as
$D_\alpha=\pi^*(D_\alpha^b)$ under the projection map
$\pi:\,\hat{X}\rightarrow B$. In addition, each point $P$, $Q$, $R$ and $S$ on the elliptic fiber $\mathcal{E}$ 
in \eqref{eq:EonBlP3} lifts to an in general rational section of the fibration $\pi:\,\hat{X}\rightarrow B$, that we denote
by $\hat{s}_P$, $\hat{s}_Q$, $\hat{s}_R$ and $\hat{s}_S$, with $\hat{s}_P$ the zero
section. The corresponding divisor classes, denoted $S_P$, $S_Q$, $S_R$ and $S_S$,  then follow from
\eqref{eq:resolvedPoints} and \eqref{eq:FullClassesuvwt} as
\beq \label{eq:divSPSQSR}
	S_P=E_3\,,\qquad S_Q=E_1\,,\qquad S_R=E_2\,,\qquad S_S=H-E_1-E_2-E_3+\cS_9+[K_B]\,,
\eeq
where we denote, by abuse of notation, the lift of the classes $H$, $E_1$, $E_2$, $E_3$ of the fiber $\text{Bl}_3\mathbb{P}^3$
in \eqref{eq:divclassBlP3} to classes in $\hat{X}$ by the same symbol.
For convenience, we collectively denote the generators of the Mordell-Weil group and their divisor classes as
\beq \label{eq:defS_m}
	\hat{s}_m=(\hat{s}_Q,\hat{s}_R,\hat{s}_S)\,,\qquad S_m=(S_Q,S_R,S_S)\,\quad m=1,2,3\,.
\eeq

The vertical divisors $D_\alpha$ together with the classes \eqref{eq:divSPSQSR} of the rational points form
a basis of $H^{(1,1)}(\hat{X})$. A basis that is better suited for applications to F-theory, however, is given by
\beq \label{eq:basisH11X}
	D_A=(\tilde{S}_P,D_\alpha,\sigma(\hat{s}_m))\,, \quad A=0,1,\ldots,h^{(1,1)}(B)+4\,,
\eeq
where the Hodge number $h^{(1,1)}(B)$  of the base $B$ counts the number of vertical divisors $D_\alpha$ in $\hat{X}$. Here
we have introduced the class \cite{Grimm:2011sk,Bonetti:2011mw}
\beq \label{eq:SPtilde}
	\tilde{S}_P=S_P+\frac{1}{2}[K_B^{-1}]\,,
\eeq
and have applied the Shioda map $\sigma$ that maps the Mordell-Weil 
group of $\hat{X}$ to a certain subspace of $H^{(1,1)}(\hat{X})$.
The map $\sigma$ is defined as
\beq \label{eq:ShiodaMapU1only}	
	\sigma(\hat{s}_m) := S_m-\tilde{S}_P-\pi(S_m\cdot\tilde{S}_P)\ ,
\eeq
where $\pi$, by abuse of notation, denotes the projection of 
$H^{(2,2)}(\hat{X})$ to the vertical homology $\pi^*H^{(1,1)}(B)$ of the
base $B$. For every $\cC$ in $H^{(2,2)}(\hat{X})$ the map $\pi$ is 
defined as
\beq \label{eq:pi}
	\pi(\mathcal{C})=(\mathcal{C}\cdot \Sigma^\alpha)D_\alpha\,,
\eeq
where we obtain the elements $\Sigma^\alpha=\pi^*(\Sigma^\alpha_b)$ in $H_{4}(\hat{X})$ as pullbacks from a dual basis  $\Sigma_b^\alpha$ to the divisors $D_\alpha^b$ 
in $B$, i.e.~$\Sigma_b^\alpha\cdot D_\beta^b=\delta^\alpha_\beta$.

Next, we list the fundamental intersections involving the divisors 
$S_P$, $S_Q$ and $S_R$ in \eqref{eq:divSPSQSR}, that will be relevant
throughout this work:
\\
\fbox{
\begin{minipage}{0.29\textwidth}
~\vspace{0.4cm}\\
\textbf{\!\!Universal intersection:\!}\!\!\\
~\vspace{0.5cm}\\
\textbf{\phantom{....\,..}Rational sections:\!}\!\!\\
~\vspace{0.6cm}\\
\textbf{Holomorphic sections:\!}\!\!\\
~\vspace{0.6cm}\\
\phantom{\ .}\textbf{\phantom{......... .}Shioda maps:\!}\!\!\\
~\vspace{0.7cm}
\end{minipage}}
\hspace{-0.1cm}\fbox{
\begin{minipage}{0.67\textwidth}
\beq\label{eq:intSecFiber}
 \,\,\,\,\phantom{....}S_P\cdot F=S_m\cdot F=1\text{ with general fiber
$F\cong \mathcal{E}$}\,, \!\!
\eeq
\bea\label{eq:SP^2}
\!\!\!\!\!\!\!\!\!\!&\!\!\!\!\pi(S_P^2+[K_B^{-1}]\cdot S_P)=\pi(S_m^2+[K_B^{-1}]\cdot S_m)=0\,,&\!\!\!\\
	\label{eq:S7S9}
	\!\!\!\!\!\!\!\!\!\!& \phantom{....} \tilde{\cS}_7=\pi(S_Q\cdot S_S)\,,\quad \hat{\cS}_7=\pi(S_R\cdot S_S)\,,\quad \cS_9=\pi(S_P\cdot S_S)\,,\nn&
\eea
\beq\label{eq:holomorphicSec}
 S_P^2+[K_B^{-1}]\cdot S_P=S_m^2+[K_B^{-1}]\cdot S_m=0\,,
\eeq
\bea \label{eq:ShiodaMapSQSR}
\sigma(\hat{s}_Q)\!\!&\!\!=\!\!&\!\!S_Q-S_P-[K_B^{-1}]\,,\nn \\
\sigma(\hat{s}_R)\!\!&\!\!=\!\!&\!\!S_R-S_P-[K_B^{-1}]\,,\\
 \sigma(\hat{s}_R)\!\!&\!\!=\!\!&\!\!S_S-S_P-[K_B^{-1}]-\cS_9\,,\nn
\eea
\end{minipage}}
~\\
The first line \eqref{eq:intSecFiber} and the second line 
\eqref{eq:SP^2} are the defining property of a section of a fibration, 
whereas the fourth line 
only holds for a holomorphic section. The third line holds because the 
collision pattern of the points  in \eqref{eq:intsSPQR} directly
translates into intersections of their divisor classes $S_m$, where we 
made use of \eqref{eq:sP3ToP2I} and \eqref{eq:sP3ToP2II}. 
In other words, \eqref{eq:SP^2} states that  divisors $\tilde{\cS}_7$, 
$\hat{\cS}_7$, $\cS_9$ are the codimension one loci where 
the sections collide with each other in the fiber $\mathcal{E}$. 
Finally, the result for
the Shioda maps of the sections follows from their definitions in 
\eqref{eq:ShiodaMapU1only} and the intersections in \eqref{eq:SP^2}.

For later reference, we also compute the intersection matrix of the Shioda maps $\sigma(\hat{s}_m)$, i.e.~the height pairing, as
\bea \label{eq:b_mnAbelian}
 \pi( \sigma(\hat{s}_m) \cdotp \sigma(\hat{s}_n) ) =\left( \begin{array}{ccc}
                    2[K_B] & [K_B] & -\cS_9+\tilde{\cS}_7 +[K_B] \\
                    \left[K_B\right] & 2[K_B] & -\cS_9+\hat{\cS}_7 + [K_B] \\
                    -\cS_9+\hat{\cS}_7 + [K_B] & -\cS_9+\hat{\cS}_7 +[K_B] & 2(-\cS_9+[K_B])  \\
                   \end{array} \right)_{mn}\,.
\eea
which readily follows from \eqref{eq:ShiodaMapSQSR} and 
\eqref{eq:SP^2}.

We note that all the above intersections \eqref{eq:intSecFiber} , \eqref{eq:SP^2}, \eqref{eq:holomorphicSec}, \eqref{eq:ShiodaMapSQSR} and 
\eqref{eq:b_mnAbelian} are in completely analogous to the ones found in
\cite{Cvetic:2012xn,Cvetic:2013nia,Cvetic:2013uta} for the case of an elliptic Calabi-Yau manifold with rank two Mordell-Weil group,
see also \cite{Park:2011ji,Morrison:2012ei,Braun:2013yti,Grimm:2013oga} for a discussion of intersections in the rank one case. 
 
\subsection{All Calabi-Yau manifolds $\hat{X}$ with $\text{Bl}_3\mathbb{P}^3$-elliptic fiber over $B$}
\label{sec:allVacua}

Finally, we are equipped to classify the generic Calabi-Yau manifolds $\hat{X}$ with elliptic fiber in $\text{Bl}_3\mathbb{P}^3$ and base
$B$. This task reduces to a classification of all possible assignments of line bundles to the sections $s_i$ in \eqref{eq:sectionsFibration} so
that the Calabi-Yau manifold $\hat{X}$ is given by the generic complete intersection \eqref{eq:EonBlP3}. Otherwise we expect additional 
singularities in $\hat{X}$, potentially
corresponding to a minimal gauge symmetry in F-theory, either from
non-toric non-Abelian singularities or from non-toric sections. 
We prove in the following that 
a generic Calabi-Yau manifold $\hat{X}$ over a base $B$ corresponds to 
a point in a certain polytope, that is related to the 
single nef-partition of the polytope of 
$\text{Bl}_3\mathbb{P}^3$ as explained below. 
The following discussion is similar in spirit to the one in 
\cite{Cvetic:2013uta,Borchmann:2013jwa}, that can agree with the toric 
classification of \cite{Braun:2013nqa}.

We begin with the basis expansion 
\beq \label{eq:DuDvDwexpansion}
	D_u=n_u^\alpha D_\alpha\,,\quad D_v=n_v^\alpha D_\alpha\,,\quad D_w=n_w^\alpha D_\alpha\,,
\eeq
into vertical divisors $D_\alpha$,
where the $n_{u}^\alpha$, $n_{v}^\alpha$ and $n_{w}^\alpha$ are integer 
coefficients. For $\hat{X}$ to be generic these coefficients are 
bounded 
by the requirement that all the sections $s_i$ in 
\eqref{eq:sectionsFibration} are generic,
i.e.~that the line bundles of which the $s_i$ are holomorphic sections 
admit holomorphic sections. This is equivalent to all divisors in 
\eqref{eq:sectionsFibration} being effective. 

First, we notice that effectiveness of the sum $[s_i]+[s_{i+10}]\geq 0$ 
in \eqref{eq:sectionsFibration} is guaranteed if the vector of integers 
$\mathbf{n}^\alpha=(n_{u}^\alpha, n_{v}^\alpha,n_{w}^\alpha)$ is an 
integral point in the rescaled polytope of $\text{Bl}_3\mathbb{P}^3$. 
Indeed, we can express the conditions of effectiveness of the divisors 
$[s_i]+[s_{i+10}]$ as the following set of inequalities in 
$\mathbb{R}^3$,
\beq \label{eq:Sums_iEffective}
	\frac{1}{-K^\alpha}\mathbf{n}^\alpha\cdot \nu_i\geq-1\,,\qquad i=1,\ldots, 7\,,
\eeq
where we also expand the canonical bundle $K_B$ of the base $B$ in terms
of the vertical divisors $D_\alpha$ as
\beq \label{eq:KBexpansion}
	[K_B]=K^\alpha D_\alpha\,
\eeq
with integer coefficients $K^\alpha$. The entries of the vectors $\nu_i$ are extracted  by first summing the rows of the two tables in 
\eqref{eq:sectionsFibration}, requiring the sum to be effective and then taking the coefficients of the the divisors $D_u$, $D_v$, $D_w$. The $\nu_i$ span the following polytope
\beq
	\Delta_{3}:=\langle\nu_i\rangle=\Big\langle\begin{pmatrix}
	-3\\
	1\\
	1
	\end{pmatrix},\begin{pmatrix}
	-1\\
	1\\
	1
	\end{pmatrix},
	\begin{pmatrix}
	-1\\
	-1\\
	1
	\end{pmatrix},
	\begin{pmatrix}
	-1\\
	1\\
	-1
	\end{pmatrix},
	\begin{pmatrix}
	1\\
	-1\\
	-1
	\end{pmatrix},
	\begin{pmatrix}
	1\\
	-1\\
	1
	\end{pmatrix}
	\begin{pmatrix}
	1\\
	1\\
	-1
	\end{pmatrix}\Big\rangle\,.
\eeq
This is precisely the dual of the polytope $\nabla_{\text{Bl}_3\mathbb{P}^3}$ of $\text{Bl}_3\mathbb{P}^3$, where the latter polytope is the convex 
hull of the following vertices,
\beq \label{eq:polySum}
\nabla_{\text{Bl}_3\mathbb{P}^3}=\Big\langle-\rho_1, \,\rho_{e_1},\, \rho_4,\, \rho_3,\, \rho_{e_2},\, \rho_{e_3},\,\rho_1 \Big\rangle\,.
\eeq 
We note that these vertices are related to  the vertices in \eqref{eq:vertsP3} and \eqref{eq:blowuprays} by an SL$(3,\mathbb{Z})$ transformation.
Thus, we confirm that the solutions to \eqref{eq:Sums_iEffective}, for which all divisors $[s_i]+[s_{i+10}]$ are 
effective, are  precisely given by vectors $\mathbf{n}^\alpha$ that 
take values for all $\alpha$ in the 
polytope of $\text{Bl}_3\mathbb{P}^3$ rescaled by the factor 
$-K^\alpha$.

Next we determine the conditions inferred from each individual class 
$[s_i]$ in \eqref{eq:sectionsFibration} being effective. We obtain the following \textit{two sets of conditions}, whose solutions, given also below, 
yield the set of all generic elliptic fibrations $\hat{X}$ with a general rank three Mordell-Weil group over a given base $B$:
\begin{center}
\fbox{
\begin{minipage}{0.9\textwidth}
\vspace{-0.4cm}
\bea \label{eq:nefPartitionVacua}
\text{1)  \phantom{......}}&\hspace{-0.5cm}0\leq([p_2]^b)^\alpha\leq -K_B^\alpha\,,&\phantom{..}\\
	\text{2)  \phantom{......}}&\hspace{-0.5cm}\mathbf{n}^\alpha\cdot \nu_i\geq K^\alpha+([p_2]^b)^\alpha\,,\quad \nu_i\in\nabla_1\,,\qquad \mathbf{n}^\alpha\cdot \nu_i\geq -([p_2]^b)^\alpha\,,\quad \nu_i\in\nabla_2\,.&\phantom{..}\nn
\eea
These conditions are solved by any $\mathbf{n}^\alpha$ being integral points in the following Minkowski sum of  the polyhedra $\nabla_1$,
$\nabla_2$ defined in \eqref{eq:nefBlP3},
\beq \label{eq:MinkowskiSum}
	\mathbf{n}^\alpha\,\, \in\,\, -(K^\alpha+([p_2]^b)^\alpha)\nabla_1+([p_2]^b)^\alpha\nabla_2\,,\qquad \forall \alpha=1,\ldots,h^{(1,1)}(B)\,.
\eeq
\end{minipage}}
\end{center}

Here the two conditions for $[p_2]^b$ in the first line of \eqref{eq:nefPartitionVacua} follow from $[s_{5}],\,[s_{18}]\geq 0$ 
and the first, respectively, second set of conditions in the second line follow from the 
first, respectively, second table in \eqref{eq:sectionsFibration}. In addition,  
we have expanded the class $[p_2]^b$ into a basis $D_\alpha$ as
\beq 
	[p_2]^b=([p_2]^b)^\alpha D_\alpha\,
\eeq
and have introduced the points $\nu_i$ that define two polytopes
\bea \label{eq:verticesNefVacua}
	\Delta_1\!&\!:=\!&\!\langle \nu_i\rangle_{0\leq i\leq 6}=\Big\langle\begin{pmatrix}
	-1\\
	0\\
	0
	\end{pmatrix},\begin{pmatrix}
	0\\
	-1\\
	0
	\end{pmatrix},
	\begin{pmatrix}
	0\\
	0\\
	-1
	\end{pmatrix},
	\begin{pmatrix}
	1\\
	-1\\
	-1
	\end{pmatrix},
	\begin{pmatrix}
	1\\
	-1\\
	0
	\end{pmatrix},
	\begin{pmatrix}
	1\\
	0\\
	-1
	\end{pmatrix}
\Big\rangle\,,\nn\\
\Delta_2\!&\!:=\!&\!\langle \nu_i\rangle_{7\leq i\leq 12}=\Big\langle\begin{pmatrix}
	-2\\
	1\\
	1
	\end{pmatrix},\begin{pmatrix}
	-1\\
	1\\
	1
	\end{pmatrix},
	\begin{pmatrix}
	-1\\
	0\\
	1
	\end{pmatrix},
	\begin{pmatrix}
	-1\\
	1\\
	0
	\end{pmatrix},
	\begin{pmatrix}
	0\\
	0\\
	1
	\end{pmatrix},
	\begin{pmatrix}
	0\\
	1\\
	0
	\end{pmatrix}
\Big\rangle\,.
\eea

Next, we show how we have constructed the solutions \eqref{eq:MinkowskiSum} to \eqref{eq:nefPartitionVacua}.
To this end, it we only have to notice that the two polytopes $\Delta_1$, $\Delta_2$ are the duals in
the sense of \eqref{eq:dualNEF} of the following two polytopes $\nabla_1$, $\nabla_2$,
\bea \label{eq:nefBlP3}
	\nabla_1=\Big\langle -\rho_1,\rho_{e_1},\rho_4,\rho_3
 \Big\rangle\,,\quad 
 \nabla_2=\Big\langle\rho_{e_2},\rho_{e_3},\rho_1
 \Big\rangle\,,
\eea
where the vectors $\rho_i$, $\rho_{e_i}$ were defined in \eqref{eq:vertsP3}, \eqref{eq:blowuprays}.
These two polytopes correspond to the unique nef-partition of \eqref{eq:polySum}.
Now, we first fix the class $[p_2]^b$
such that the first conditions in \eqref{eq:nefPartitionVacua} are met. Second, for each allowed class
for $[p_2]^b$ we solve the second set of conditions in \eqref{eq:nefPartitionVacua} for the vectors 
$\mathbf{n}^\alpha$. However, these are just the duality relations between the $\Delta_i$ and $\nabla_j$, rescaled
by appropriate factors. Consequently, the solutions are precisely given by the 
integral points in the Minkowski sum of the polyhedra in \eqref{eq:MinkowskiSum}.
Here we emphasize again that both coefficients in \eqref{eq:MinkowskiSum}
are positive integers by means of the first condition in \eqref{eq:nefPartitionVacua}. 

In summary, we have shown that for a given base $B$ a generic elliptically fibered Calabi-Yau manifold $\hat{X}$ 
with general elliptic fiber $\mathcal{E}$ given by \eqref{eq:EonBlP3} in $\text{Bl}_3\mathbb{P}^3$ 
corresponds to an integral point $\mathbf{n}^\alpha$
in the polyhedron \eqref{eq:MinkowskiSum} for every $\alpha$ and for every class $[p_2]^b$ obeying 
$0\leq [p_2]^b\leq [K_B^{-1}]$. The coordinates of the point $\mathbf{n}^\alpha$ are the coefficients
of the divisors $D_u$, $D_v$, $D_w$ in the expansion \eqref{eq:DuDvDwexpansion} into vertical divisors
$D_\alpha$.

\section{Matter in F-Theory Compactifications with a Rank Three Mordell-Weil Group}
\label{sec:Codim2Sings}

In this section we analyze the codimension two singularities
of the elliptic fibration of $\hat{X}$ to determine the matter representations of corresponding F-theory 
compactifications to six and four dimensions. We find 14 different singlet representations in 
sections \ref{sec:singsOfRationalSections} and \ref{sec:SingsWSF}. Then, we determine the explicit matter multiplicities of these 14 matter fields in six-dimensional F-theory compactification on a Calabi-Yau threefold
$\hat{X}_3$ with a general two-dimensional base $B$ in section \ref{sec:MatterMultiplicities}. The following discussion 
is based on techniques developed in  \cite{Cvetic:2013nia,Cvetic:2013uta,Cvetic:2013jta}  
for the case of a rank two Mordell-Weil group, to which we refer for more background on  some technical details.

We begin with an outline of the general strategy to determine matter in an F-theory compactification on a 
Calabi-Yau manifold with a higher rank Mordell-Weil group.  First, we recall that in general rational curves $c_{\text{mat}}$ obtained from resolving a singularity of the elliptic fibration at codimension two in the base $B$ give rise to matter in F-theory 
due to the presence of light M2-brane states in the F-theory limit. In elliptically fibered Calabi-Yau manifolds with a 
non-Abelian gauge symmetry in F-theory, these codimension two singularities are located on the divisor in the base $B$, 
which supports the 7-branes giving rise to the non-Abelian gauge group. Technically, the discriminant
of the elliptic fibration takes the form $\Delta= z^n(k+\mathcal{O}(z))$, where $z$ vanishes along the 7-brane
divisor and $k$ is a polynomial independent of $z$. Then, the codimension two singularities are precisely given by the 
intersections of $z=0$ and $k=0$.

This is in  contrast to elliptic fibrations with only a non-trivial Mordell-Weil group, i.e.~only an Abelian gauge 
group, since the elliptic fibration over codimension one has only $I_1$-singularities and the discriminant does
not factorize in an obvious way. Thus, the codimension two codimension singularities are not 
contained in a simple divisor in $B$ and have to be studied directly.  
In fact, the existence of a rational section, denoted by say $\hat{s}_Q$, 
means that there is a solution to the Weierstrass form (WSF) of the form 
$[x^Q:y^Q:z^Q]=[g^Q_{2}:g^Q_{3}:1]$.\footnote{Sections with $z^Q=b$ for a section $b$ of a line bundle 
$\mathcal{O}([b])$ 
on the base $B$ and with $g^Q_2$,  $g^Q_3$ sections of $K_B^{-2}\otimes\mathcal{O}(2[b])$, respectively, 
$K_B^{-3}\otimes\mathcal{O}(3[b])$, can be studied similarly. We only have to assume that we are at a locus with 
$b\neq 0$. Then we can employ the $\mathbb{C}^*$-action to set $z^Q=1$, $x^Q=\frac{g^Q_2}{b^2}$, 
$y^Q=\frac{g^Q_3}{b^3}$.} Here $g^Q_2$ and $g_3^Q$ are sections of 
$K_B^{-2}$ and $K_B^{-3}$, respectively.\footnote{For concreteness and for 
comparison to \cite{Morrison:2012ei,Cvetic:2013nia}, in the special case of 
the base $B=\mathbb{P}^2$, the sections 
$g^Q_2=g_{6}$, $g^Q_3=g_{9}$ are polynomials of degree $6$, respectively, $9$}. Thus, the presence of 
$\hat{s}_Q$ implies the factorization
\beq \label{eq:WSF}
(y-g^Q_{3}z^3)(y+g^Q_{3}z^3)=(x-g^Q_{2}z^2)(x^2+g^Q_{2}xz^2+g^Q_{4}z^4)\,
\eeq
for appropriate $g^Q_4$. Parametrizing the discriminant $\Delta$ in terms of the polynomials in \eqref{eq:WSF}, we 
see that it vanishes of order two at the codimension two loci in $B$ reading
\beq \label{eq:conifold}
g^Q_{3}=0\,, \qquad  \qquad \hat{g}^Q_{4}:=g^Q_{4}+2(g^Q_{2})^2=0\,.
\eeq 
These two conditions lead to a factorization of both sides of \eqref{eq:WSF}, so that a conifold singularity 
is developed at $y=(x-g^Q_2z^2)=0$. 

It is evident that the section $\hat{s}_Q$ passes automatically through the singular point of the elliptic curve. 
Thus,  in the resolved elliptic curve $\mathcal{E}$ where the singular point $y=(x-g^Q_2z^2)=0$ is replaced by
a Hirzebruch-Jung sphere tree of intersecting $\mathbb{P}^1$'s,\footnote{In F-theory compactifications with only
Abelian groups the resolved elliptic fibers are expected to be $I_2$-curves, i.e.~two $\mathbb{P}^1$'s intersecting at 
two points.} the section $\hat{s}_Q$ automatically intersects at least one $\mathbb{P}^1$. This implies that the loci 
\eqref{eq:conifold} in the base contain matter charged under U$(1)_Q$ associated to $\hat{s}_Q$, as can be seen from the charge formula 
\beq \label{eq:U1charge}
	q_Q=c_{\text{mat}}\cdot (S_Q-S_P)\,.
\eeq
Here $S_Q$, $S_P$ denote the divisor classes of 
$\hat{s}_Q$ and the zero section $\hat{s}_P$, respectively. 
In fact, the locus \eqref{eq:conifold} contains the codimension two loci supporting \textit{all}  matter
charged under U$(1)_Q$, without distinguishing 
between matter with different U$(1)_Q$-charges. The loci of the different matter representations correspond to the
irreducible components of \eqref{eq:conifold}, that can in principle be obtained by finding all 
associated prime ideals of \eqref{eq:conifold} of codimension two in $B$. Unfortunately, in many concrete setups this is computationally unfeasible 
and we have to pursue a different strategy to obtain the individual matter representations that has already been successful in the rank two case in 
\cite{Morrison:2012ei,Cvetic:2013nia}.

For the following analysis of codimension two singularities of $\hat{X}$ we identify the irreducible components of \eqref{eq:conifold} 
corresponding to different matter representations in two qualitatively different ways:
\begin{itemize}
\item[1)] One type of codimension two singularities corresponds to singularities of the sections $\hat{s}_m$ and $\hat{s}_P$. This analysis, see section \ref{sec:singsOfRationalSections}, is performed in the presentation of $\mathcal{E}$ as the complete intersection \eqref{eq:EonBlP3} in 
$\text{Bl}_3\mathbb{P}^3$, where the rational sections are given by \eqref{eq:PQRSMinors}. In fact, 
when a rational section $\hat{s}_m$ or the zero section  $\hat{s}_P$ is ill-defined, the resolved elliptic curve 
splits into an $I_2$-curve with one $\mathbb{P}^1$ representing the original singular fiber and the other $\mathbb{P}^1$ 
representing the singular section.
\item[2)] The second type of codimension two singularities has to be found directly in the Weierstrass model. The basic
idea is isolate special solutions to \eqref{eq:conifold} by supplementing the two equations \eqref{eq:conifold} by further constraints 
that have to vanish in addition in order for a certain matter representation to be present. We refer to section \ref{sec:SingsWSF} for 
concrete examples. It is then
possible to find the codimension two locus along which all these constraints vanish simultaneously. We note that
for the geometry $\hat{X}$ there are three rational sections, thus, three factorizations of the form
\eqref{eq:WSF} and loci \eqref{eq:conifold}, that have to be analyzed separately. 
\end{itemize}

A complete analysis of codimension two singularities following the above two-step strategy should 
achieve a complete decomposition of \eqref{eq:conifold} for all sections of $\hat{X}$ into irreducible components. 
It would be interesting to prove this mathematical for the codimension two singularities of $\hat{X}$ we find in this 
section. As a consistency check of our analysis of codimension two singularities we find, we determine the full spectrum, including 
multiplicities, of charged hypermultiplets of a six-dimensional F-theory compactification and check that six-dimensional 
anomalies are cancelled, cf.~section \ref{sec:MatterMultiplicities}.

\subsection{Matter at the Singularity Loci of Rational Sections}
\label{sec:singsOfRationalSections}

Now that the strategy is clear, we will look for the first type of singularities in this subsection. These are the codimension two loci in the base 
where the rational sections are singular in $\text{Bl}_3\mathbb{P}^3$. This  precisely happens when the coordinates \eqref{eq:resolvedPoints}, 
\eqref{eq:PQRSMinors} of any of the rational sections  take values in the Stanley-Reisner ideal \eqref{eq:SRBlP3} of $\text{Bl}_3\mathbb{P}^3$. 

There are two reasons why codimension two loci with singular rational sections are good candidates for $I_2$-fibers. First, the elliptic 
fibration of $\hat{X}$ is smooth\footnote{This is clear for toric bases $B$.}, thus, the indeterminacy of the coordinates of the sections
in the fiber may imply that the section is not a point, but an entire $\mathbb{P}^1$. Second, as was remarked in \cite{Morrison:2012ei} and 
\cite{Cvetic:2013nia}, if we approach the codimension two singularity  of the section along a line in the base $B$ the section has a well defined 
coordinate given by the slope of the line. Thus, approaching the singularity along lines of all possible slopes
the section at the singular point is identified with the $\mathbb{P}^1$ formed by all slopes.  In fact, specializing the elliptic curve
to each locus yielding a singularity of a rational section we observe a splitting of the elliptic curve into an $I_2$-curve.
We note that it is crucial to work in $\text{Bl}_3\mathbb{P}^3$, because only in this space the fiber is fully resolved space by the 
exceptional divisors $E_i$, in contrast to the curve \eqref{eq:1st} in $\mathbb{P}^3$.

\subsubsection{The vanishing of two minors: special singularities of $\hat{s}_S$}
\label{sec:TwoMinors}

In order to identify singularities of rational sections, let us take a close look at the Stanley-Reisner ideal \eqref{eq:SRBlP3}. It contains monomials 
with two variables of the type $e_i e_j$ and monomials with three variables of the type $uXY$, where $X$ and $Y$ are two variables out of the set $\{v,w,t\}$.  In this subsection we look for singular sections whose coordinates are forbidden by the elements $e_ie_j$. 

From the coordinates \eqref{eq:PQRSMinors} of the rational sections we infer that this type of singular behavior can only occur for the section $\hat{s}_S$, whose coordinates in the fiber $\mathcal{E}$ are
\beq \label{eq:Sfiber}
S=[0:1:1:1 :s_{19} s_{8} -  s_{18} s_9: s_{10} s_{18} -  s_{20} s_8:s_{10} s_{19} - s_{20} s_9]\,. \\
\eeq
There are three codimension two loci where $S$ is singular, reading
\beq \label{eq:TwoMinorsLoci}
\{s_8=s_{18}=0\}\,, \qquad  \{s_9=s_{19}=0\}\,, \qquad \{s_{10}=s_{20}=0\}\,.
\eeq  
It is important to note that the matrices \eqref{eq:MS}, \eqref{eq:MatsPQR} retain rank two at these loci, since only two of their 
$2\times 2$-minors, being identified with the coordinates \eqref{eq:PQRSMinors}, have vanishing determinant.
Next, we inspect the constraint \eqref{eq:EonBlP3} of the elliptic curve at these loci.

At all these three codimension two loci, we see that the elliptic curve in \eqref{eq:EonBlP3} takes the common form
\beq \label{eq:CISing}
  Au + B Y= 0\,, \quad Cu + D Y= 0\,.
\eeq
Here $Y$ is one of the variables $\{v,w,t\}$ and the polynomials $B$, $D$ are chosen to be independent of $u$ and $Y$, which fixes the 
polynomials $A$, $C$ uniquely.  This complete intersection describes a reducible curve. 
This can be seen by rewriting it as 
\bea  \label{eq:I2curveSimple}
(AD-BC) u = 0 \,, \qquad Au + B Y=Cu + D Y= 0\,,
\eea
which we obtained by solving for the variable $Y$ in the first equation of \eqref{eq:CISing} and requiring consistency with the second equation. 

Now, we directly see that one solution to \eqref{eq:I2curveSimple} is given by $\{u=0,\,Y=0\}$.  This  is 
a $\mathbb{P}^1$ as is clear from the remaining generators of the SR-ideal after setting the coordinates that are not allowed to vanish to 
one using the $\mathbb{C}^*$-actions. The second solution, which also describes a $\mathbb{P}^1$, is given by the vanishing of 
the determinant in the first equation in \eqref{eq:I2curveSimple}, which implies that the two constraint in the second equation become 
dependent.
Thus, the two $\mathbb{P}^1$'s of the $I_2$-curve are given by
\beq \label{eq:P1sTwoMinors}
c_1=\{u=0,\,Y=0\}\,, \qquad   \quad c_2=\{AD-BC=0,\, Cu + D Y=0\}\,.
\eeq

As an example, let us look at the loci $\{s_8=s_{18}=0\}$ in \eqref{eq:TwoMinorsLoci} in detail. In this case the elliptic curve $\mathcal{E}$ given 
in \eqref{eq:EonBlP3} takes the form 
\bea 
&u(s_2e_1 e_2 e_3 u+s_5e_1 e_2  t+s_{6}e_2 e_3 v  +s_7e_1 e_3  w)  = t (s_9e_2  v +s_{10}e_1  w)   \,, & \\
&u(s_{12}e_1 e_2 e_3 u + s_{15}e_1 e_2  t + s_{16}e_2 e_3 v + s_{17}e_1 e_3  w )  = t (s_{19}e_2  v + s_{20}e_1  w)\,. & \nn
\eea
This complete intersection is in the form \eqref{eq:CISing} by identifying $Y=t$ and  setting
\bea
&A=(s_2e_1 e_2 e_3 u+s_5e_1 e_2  t+s_{6}e_2 e_3 v  +s_7e_1 e_3  w)\,, \qquad B=-(s_9e_2  v +s_{10}e_1  w)\,,&  \\
&C=(s_{12}e_1 e_2 e_3 u + s_{15}e_1 e_2  t + s_{16}e_2 e_3 v + s_{17}e_1 e_3  w ),\, \qquad D=-(s_{19}e_2  v + s_{20}e_1  w)\,.& \nn
\eea 
Then the two $\mathbb{P}^1$'s of the $I_2$-curve are given by $c_1$, $c_2$ in \eqref{eq:P1sTwoMinors}. 

Equipped with the equations for the individual curves $c_1$, $c_2$ we can now  calculate the intersections with the sections and the charge of the 
hypermultiplet that is supported there. The intersections of the curve defined $c_1$ can be readily obtained from the toric intersections of 
$\text{Bl}_3\mathbb{P}^3$. It has intersection $-1$ with the section $S_S$, intersection one with the sections $S_Q$, $S_R$ and zero with $S_P$, 
where the last intersection is clear from the existence of the term $e_3 t$ in the Stanley-Reisner ideal \eqref{eq:SRBlP3}.  The intersections
with $c_2$ can be calculated either directly from \eqref{eq:P1sTwoMinors} or from the fact, that the intersections of a section 
with the total class $F=c_1+c_2$ have to be one.

We summarize our findings as:
\beq
\text{
\begin{tabular}{|c|c|c|c|c|c|} \hline
Loci & Curve & $\cdot S_P$ & $\cdot S_Q$ & $\cdot S_R$ & $\cdot S_S$ \rule{0pt}{14pt}\\ \hline
$s_{8}=s_{18}=0$ \rule{0pt}{12pt}& $c_1=\{u=t=0\}$ & $0$ & $1$ & $1$ & $-1$ \\ 
 & $c_2$ & 1 &0 & 0 & 2\\ \hline
$s_{9}=s_{19}=0$\rule{0pt}{12pt} & $c_1=\{u=w=0\}$ & $1$ & 1 & $0$ & $-1$ \\
 & $c_2$ & 0 &0 & 1 & 2\\ \hline
$s_{10}=s_{20}=0$\rule{0pt}{12pt} & $c_1=\{u=v=0\}$ & $1$ & 0 & 1 & $-1$ \\
 & $c_2$ & 0 &1 & 0 & 2\\ 
 \hline
\end{tabular}
}
\eeq
Here we denoted the intersection pairing by `$\cdot$' and we also computed the intersections of the sections with the  $I_2$-curves at the other 
two codimension two loci in \eqref{eq:TwoMinorsLoci}. In these cases, we identified $Y=w$, respectively, $Y=v$.

We proceed with the calculation of the charges in each case employing the charge formula \eqref{eq:U1charge}. We note that the isolated curve $c_{mat}$ is always the curve in the $I_2$-fiber that that does not intersect the zero section $S_P$. We obtain the charges:
\beq \label{eq:TwoMinorsCharges}
\text{
\begin{tabular}{|c|c|c|c|c|c|} \hline
Loci &  $q_Q$ & $q_R$ & $q_S$ \rule{0pt}{12pt} \\ \hline
$s_{8}=s_{18}=0$ & 1 & $1$ & $-1$ \\ 
$s_{9}=s_{19}=0$ &  0 & $1$ & $2$ \\
$s_{10}=s_{20}=0$ &  1 & 0 & $2$ \\ \hline
\end{tabular}
}
\eeq

\subsubsection{The vanishing of three minors: singularities of all sections}
\label{sec:ThreeMinors}

The remaining singularities of the rational sections occur if the three of the determinants of the minors of the matrices \eqref{eq:MS}, 
\eqref{eq:MatsPQR} vanish. This implies that three coordinates \eqref{eq:PQRSMinors} of a section are forbidden by the 
SR-ideal \eqref{eq:SRBlP3}, which happens also for  the sections 
$\hat{s}_P$,  $\hat{s}_Q$, $\hat{s}_R$,  in addition to $\hat{s}_S$, due to the elements $uXY$ with  $X$, $Y$ in $\{v,w,t\}$. 

Before analyzing these loci, we emphasize that the three vanishing conditions are a codimension two 
phenomenon because the vanishing of the determinants of three minors of the same matrix is not independent. In fact, these 
codimension two loci can be viewed as determinantal varieties describing 
the loci where  the rank of each of the matrices in \eqref{eq:MS}, \eqref{eq:MatsPQR} jump from two to one, which is clearly a codimension
two phenomenon.

Concretely, for the section $\hat{s}_P$ to be singular, the three minors that have to vanish are $|M^P_3|=|M^P_2|=|M^P_1|=0$, which implies the 
conditions
\beq \label{eq:LociPS}
  \frac{s_5}{s_{15}}=\frac{s_{10}}{s_{20}}=\frac{s_9}{s_{19}}\,.
 \eeq
Similarly, for $\hat{s}_Q$ to be singular, we impose $|M^Q_3|=|M^Q_2|=|M^Q_1|=0$, which yields
 \beq \label{eq:LociQS}
 \frac{s_6}{s_{16}}=\frac{s_8}{s_{18}}=\frac{s_9}{s_{19}}\,.
\eeq
For a singular section $\hat{s}_R$, we require $|M^R_3|=|M^R_2|=|M^R_1|=0$, which is equivalent to
\beq \label{eq:LociRS}
 \frac{s_{10}}{s_{20}}=\frac{s_8}{s_{18}}=\frac{s_7}{s_{17}}\,.
\eeq
Finally,  the  section $\hat{s}_S$ is singular at $|M^Q_3|=|M^R_3|=|M^P_3|=0$, or equivalently at
\beq \label{eq:LociSS}
 \frac{s_{10}}{s_{20}}=\frac{s_8}{s_{18}}=\frac{s_9}{s_{19}}\,.
\eeq
We remark that the vanishing of the three minors in all these cases excludes the loci \eqref{eq:TwoMinorsLoci} of the previous subsection.

All these singularities imply a reducible curve of a form similar to \eqref{eq:DegenerationOfQuadrics}, however, adapted to the ambient space  
$\text{Bl}_3\mathbb{P}^3$. In fact, at each of the loci \eqref{eq:LociPS}-\eqref{eq:LociSS} the complete intersection \eqref{eq:EonBlP3} takes the 
form
\beq \label{eq:AXCX}
A X +  B Y =0\,, \qquad
C X + D Y =0 \,,
\eeq
for appropriate polynomials $A$, $B$, $C$, $D$ with $A$ and $C$ collinear, that is $A= aC$, and the pair of coordinates $[X:Y]$ forming a 
$\mathbb{P}^1$.\footnote{When $\hat{s}_S$ becomes 
singular, we identify $Y=u$ and $X=1$.  However, $A$, $C$ still become collinear and the argument applies.} 
Then, we can multiply the second equation by $a$ and subtract from the first equation, to obtain
\beq \label{eq:I2curvedP2}
(B-aD) Y =0\,, \qquad
A X + B Y=0\,.
\eeq
From this we see that the two solutions are given by 
\beq \label{eq:c1c2curves}
c_1=\{Y=A=0\} \,, \qquad c_2=\{B-aD=AX+BY=0\}\,,
\eeq
that describe two $\mathbb{P}^1$'s intersecting at two points.
Thus  the complete intersection
\eqref{eq:I2curvedP2} is an $I_2$-curve.

\subsubsection*{One example in detail}

Let us focus on the locus in  \eqref{eq:LociQS} where the section $\hat{s}_Q$ is singular. The complete intersection  \eqref{eq:EonBlP3} then takes 
the form
\bea
&v (- e_2 s_9 t  + e_2 e_3 s_6 u  + e_3 s_8  w )+e_1 ( e_2 s_5 t u + e_2 e_3 s_2 u^2  - s_{10} t w + s_7 e_3 u w ) =0 \,, \nn &\\
&v (- e_2 s_{19} t  + e_2 e_3 s_{16} u   + e_3 s_{18}  w )+e_1 ( e_2 s_{15} t u + e_2 e_3 s_{12} u^2 - s_{20} t w + e_3 s_{17} u w )  =0\,.& \nn
\eea
This is of the form \eqref{eq:AXCX}  as we see by identifying $X=v$ and $Y=e_1$  and by setting
\bea \label{eq:ABCDconcrete}
&A=- e_2 s_9 t  + e_2 e_3 s_6 u  + e_3 s_8  w \,, \quad  B= e_2 s_5 t u + e_2 e_3 s_2 u^2  - s_{10} t w + s_7 e_3 u w\,,&\\
&C=- e_2 s_{19} t  + e_2 e_3 s_{16} u   + e_3 s_{18}  w  \,,\quad  D=e_2 s_{15} t u + e_2 e_3 s_{12} u^2 - s_{20} t w + e_3 s_{17} u w\,& \nn
\eea
with $A=(s_8/s_{18})C$ collinear at the locus \eqref{eq:LociQS} . Then, the two $\mathbb{P}^1$'s in this $I_2$-curve are given by
\eqref{eq:c1c2curves} with the identifications \eqref{eq:ABCDconcrete}.

Next, we obtain the intersections of the  curves $c_1$, $c_2$ with the rational sections, that follow directly from the toric intersections of 
$\text{Bl}_3\mathbb{P}^3$. We find the intersections
\beq
\text{
\begin{tabular}{|c|c|c|c|c|c|} \hline
Loci & Curve & $\cdot S_P$ & $\cdot S_Q$ & $\cdot S_R$ & $\cdot S_S$ \rule{0pt}{14pt} \\ \hline
$|M^Q_3|=|M^Q_2|=|M^Q_1|=0$ \rule{0pt}{14pt} & $c_1$ & 0& $-1$ & 0 & 1 \\
 & $c_2$ & 1& $2$ & 1 & 0 \\ \hline
\end{tabular}
}
\eeq
As expected, the total fiber $F=c_1+c_2$ has intersections  $S_m \cdot F=1$ with all sections.

Repeating the procedure with the other codimension two loci \eqref{eq:LociPS}, \eqref{eq:LociRS} and \eqref{eq:LociSS}, we obtain the 
intersections of the split elliptic curve with the sections as
\beq \label{eq:dP2Int}
\text{ 
\begin{tabular}{|c|c|c|c|c|c|} \hline
Loci & Curve & $\cdot S_P$ & $\cdot S_Q$ & $\cdot S_R$ & $\cdot S_S$\rule{0pt}{14pt} \\ \hline
$|M^R_3|=|M^R_2|=|M^R_1|=0$\rule{0pt}{14pt} & $c_1$ & 0& 0 & $-1$ & 1 \\
 & $c_2$ & 1 & 1 & 2 & 0 \\ \hline 
$|M^P_3|=|M^P_2|=|M^P_1|=0$\rule{0pt}{14pt}& $c_1$ & $-1$& 0 & $0$ & 1 \\
 & $c_2$ & 2 & 1 & 1 & 0 \\ \hline 
$|M^Q_3|=|M^R_3|=|M^P_3|=0$ \rule{0pt}{14pt}& $c_1$ & $1$& 1 & 1 & $-1$ \\ 
 & $c_2$ & 0 & 0 & 0 & 2 \\ \hline 
\end{tabular}
}
\eeq
With these intersection numbers and the charge formula \eqref{eq:U1charge} we obtain the charges
\beq \label{eq:ThreeMinorCharges}
\text{
\begin{tabular}{|c|c|c|c|} \hline
Loci  & $q_Q$ & $q_R$ & $q_S$ \rule{0pt}{14pt} \\ \hline
$|M^Q_3|=|M^Q_2|=|M^Q_1|=0$ \rule{0pt}{14pt}&  $-1$ & 0 & 1 \\ 
$|M^R_3|=|M^R_2|=|M^R_1|=0$ \rule{0pt}{14pt}&  0 & $-1$ & 1 \\ 
$|M^P_3|=|M^P_2|=|M^P_1|=0$\rule{0pt}{14pt}& $-1$ & $-1$ & $-2$ \\
$|M^Q_3|=|M^R_3|=|M^P_3|=0$ \rule{0pt}{14pt}& 0 & 0 & $2$ \\ \hline
\end{tabular}
}
\eeq

\subsubsection*{Relation to $dP_2$}

In section \ref{sec:QuadsToCubicResolved} we saw that the elliptic curve $\mathcal{E}$ can be mapped to two\footnote{There are actually three
$dP_2$ maps if we are willing to give up the zero point as a toric point. See section \ref{sec:QuadsToCubicResolved} for more details.} non-generic 
anti-canonical hypersurfaces in $dP_2$. It is expected that some of the singularities we just found map to the singularities in the $dP_2$-elliptic 
curve. We recall from  \cite{Cvetic:2013nia,Borchmann:2013jwa}, that the Calabi-Yau hypersurfaces \eqref{eq:cubicP2}, \eqref{eq:cubicP2_v} in 
$dP_2$  have singular sections at the codimension two loci given by $\tilde{s}_3=\tilde{s}_7=0$ ($\hat{s}_3=\hat{s}_7=0$), $\tilde{s}_8=\tilde{s}_9=0$ ($\hat{s}_8=\hat{s}_9=0$) and $\tilde{s}_7=\tilde{s}_9=0$ ($\hat{s}_7=\hat{s}_9=0$), respectively.

In tables \eqref{eq:sP3ToP2I} and \eqref{eq:sP3ToP2II} we readily identified the minors of the matrices in \eqref{eq:MatsPQR} with the some of the 
coefficients $\tilde{s}_i$ and $\hat{s}_j$. This implies a relationship between the singular codimension two loci of the elliptic curves in $\text{Bl}_3\mathbb{P}^3$ and in the two $dP_2$-varieties, that we summarize  in the following table:
\beq
\text{ 
\begin{tabular}{|c|c|c|} \hline
$\text{Bl}_3\mathbb{P}^3$-singularity & Singularity of & Singularity  of\rule{0pt}{14pt}\\
  & curve in \eqref{eq:cubicP2} & curve in \eqref{eq:cubicP2_v}\\ \hline
$|M^Q_3|=|M^Q_2|=|M^Q_1|=0$ \rule{0pt}{14pt}& $\tilde{s}_3=\tilde{s}_7=0$ & $Q$ non-toric  \\
$|M^R_3|=|M^R_2|=|M^R_1|=0$ \rule{0pt}{14pt}& $R$ non-toric & $\hat{s}_3=\hat{s}_7=0$ \\
$|M^P_3|=|M^P_2|=|M^P_1|=0$ \rule{0pt}{14pt}& $\tilde{s}_8=\tilde{s}_9=0$ & $\hat{s}_8=\hat{s}_9=0$ \\
$|M^Q_3|=|M^R_3|=|M^P_3|=0$ \rule{0pt}{14pt}& $\tilde{s}_7=\tilde{s}_9=0$ & $\hat{s}_7=\hat{s}_9=0$ \\ \hline
\end{tabular}
}
\eeq
In each case, three out of the four singular loci \eqref{eq:ThreeMinorCharges} yield singularities of the toric sections in the $dP_2$-elliptic curve. The other singular locus in the curve in $\text{Bl}_3\mathbb{P}^3$ is not simply given by the vanishing of two coefficients $\tilde{s}_i$, respectively $\hat{s}_j$, because the non-toric rational sections becomes singular. Nevertheless, the elliptic curve in $dP_2$ admits a
factorization at the singular locus of the non-toric section, i.e.~it splits into an $I_2$-curve, due to the non-genericity of the corresponding coefficients $\tilde{s}_i$ or $\hat{s}_j$.

\subsection{Matter from Singularities in the Weierstrass Model}
\label{sec:SingsWSF}

As mentioned in the introduction of this subsection, \textit{all} the loci of matter charged under a section $\hat{s}_m$ satisfy the equations $g_{3}^m=0$ and $\hat{g}^m_{4}=0$. Since we have three rational sections $\hat{s}_m$, the WSF admits three 
possible factorizations  of the form \eqref{eq:WSF}, each of which implying a singular elliptic fiber at the loci  $g^{Q,R,S}_3=\hat{g}^{Q,R,S}_{4}=0$ 
with $\hat{g}^{R,S}_{4}$ defined analogous to \eqref{eq:conifold}.
In this subsection we separate solutions to these equations by requiring additional constraints to vanish.

We can isolate matter with simultaneous U(1)-charges. The idea is the following. If the matter is charged under two sections, both sections have to 
pass through the singularity in the WSF. This requires the $x$-coordinates $g_2^{m_1}$, $g_2^{m_2}$ of the sections to agree\footnote{Here we 
assume that the $z-$-coordinates of both sections are $z=1$, for simplicity.},
\beq \label{eq:deltaG2}
 \delta g_2^{m_1,m_2} := g_2^{m_1}- g_2^{m_2}\stackrel{!}{=}0,
\eeq
for any two sections $\hat{s}_{m_1}$ and $\hat{s}_{m_2}$.
The polynomial \eqref{eq:deltaG2} has a smaller degree than the other two conditions \eqref{eq:conifold} and in fact it will be one of the two 
polynomials of the complete intersection describing the codimension two locus. The other constraint will be $g_3^{m}=0$ for $m$ either $m_1$ or 
$m_2$.

If we solve for two coefficients in these two polynomials and insert the solution back into the elliptic curve \eqref{eq:EonBlP3} we observe a 
reducible curve of the form \eqref{eq:I2curvedP2}. In this $I_2$-curve, one  $\mathbb{P}^1$ is automatically intersected once by both sections 
$\hat{s}_{m_1}$ and $\hat{s}_{m_2}$. This means that a generic solution of equations \eqref{eq:conifold}, \eqref{eq:deltaG2} support matter with 
charges one under U$(1)_{m_1}\times$U$(1)_{m_2}$.

Let us be more specific for matter charged under the sections $\hat{s}_Q$ and $\hat{s}_{R}$, that is matter transforming under 
U$(1)_Q\times$U$(1)_R$. The conditions \eqref{eq:conifold} and \eqref{eq:deltaG2}  read
\beq \label{eq:collisionQR}
	\delta g_2^{QR}:=g_2^Q-g_2^R\stackrel{!}{=}0\,, \qquad g^Q_3=0\,, \qquad \hat{g}^Q_{4}=0\,,
\eeq
and the codimension to locus is given by the complete intersection $\delta g_2^{QR}=g^Q_3=0$. In fact the constraint  $\hat{g}^Q_{4}$, $\hat{g}^R_{4}$ are in the ideal generate by $\langle \delta g_2^{QR}, g^Q_3\rangle$.

We proceed to look for matter charged under U$(1)_Q\times$U$(1)_S$. In this case, because of the  section $\hat{s}_S$ having a non-trivial 
$z$-component, the right patch of the WSF is $z\equiv\tilde{z}^S=s_{10}s_{19}-s_{20}s_9$, c.f. \eqref{eq:QRSgeneralForm}. Thus, the constrains 
\eqref{eq:conifold} and \eqref{eq:deltaG2} take the form
\beq
\delta g_2^{QS} := g_2^{S} - (\tilde{z}^S)^2 g_2^{Q}\stackrel{!}{=}0\,, \qquad g^S_3=0\,, \qquad \hat{g}^S_{4}=0\,.
\eeq
Instead of using these polynomials, we will use two slightly modified polynomials that generate the same ideal.  They were defined in 
\cite{Cvetic:2013nia} where they were denoted by $\delta g^\prime_6$ and $g^\prime_9$ and defined as
\beq  \label{eq:modifiedQSLoci}
\delta (g_2^{QS})^\prime := \tilde{s}_7 \tilde{s}_8^2+\tilde{s}_9(-\tilde{s}_6 \tilde{s}_8+\tilde{s}_5 \tilde{s}_9)=0, \qquad ( g_3^{QS} )^\prime := \tilde{s}_3 \tilde{s}_8^2-\tilde{s}_2 \tilde{s}_8 \tilde{s}_9+\tilde{s}_1 \tilde{s}_9^2=0\,,
\eeq
Here we have to use the map \eqref{eq:sP3ToP2I} to obtain these polynomials in terms of the coefficients $s_i$.
We will see in section \ref{sec:MatterMultiplicities} that these polynomials are crucial to obtain the matter multiplicities of this type of charged 
matter fields.

Similarly, for matter charged under U$(1)_R\times$U$(1)_S$ we demand
\beq
\delta g_2^{RS} := g_2^{S} - (\tilde{z}^S)^2 g_2^{R}\stackrel{!}{=}0\,, \qquad g^S_3=0\,, \qquad \hat{g}^S_{4}=0\,.
\eeq
For this type of locus we will also use the modified polynomials $\delta (g_2^{RS})^\prime$ and $\delta (g_3^{RS})^\prime$ that can be obtained 
from \eqref{eq:modifiedQSLoci} by replacing all the coefficients $\tilde{s}_i\rightarrow\hat{s}_i$ and by using \eqref{eq:sP3ToP2II}.

Next, we look for matter charged under all U(1) factors U$(1)_Q\times$U$(1)_R\times$U$(1)_S$. This requires the three sections to collide and pass through the singular point $y=0$ in the WSF,  at codimension two. The four polynomials that are required to vanish simultaneously are
\beq \label{eq:QRSLoci}
 \delta  g^{QS}_2=0\,, \qquad (\tilde{z}^S)^2 \delta g^{RS}_2=0\,, \qquad  g^{S}_3=0\,, \qquad \hat{g}^{S}_{4}=0\,, 
\eeq
where the first two conditions enforce a collision of  the three sections in the elliptic fiber. In order for a codimension two locus to satisfy all these 
constraints simultaneously, all the polynomials \eqref{eq:QRSLoci} should factor as
\beq
p = h_1 p_1 + h_2 p_2\,,
\eeq
where $h_1$ and $h_2$ are the polynomials whose zero-locus defines the codimension two locus in question. To obtain the polynomials we use 
the Euclidean algorithm twice. We first divide all polynomials in \eqref{eq:QRSLoci} by the lowest order polynomial available, which is 
$\delta g_2^{QR}$ and take the biggest common factor from all residues. This is the polynomial $h_1$ and it reads
\bea \label{eq:h1}
h_1&=&(s_{10}^2 s_{15} s_{16} s_{19} + s_{10}^2 s_{12} s_{19}^2 + s_{10} s_{15} s_{18} s_{19} s_{5} + 
  s_{10} s_{17} s_{19}^2 s_{5}  - s_{10} s_{16} s_{19} s_{20} s_{5} \nn \\ && - s_{18} s_{19} s_{20} s_{5}^2 - 
  s_{10} s_{15}^2 s_{18} s_{9} - s_{10} s_{15} s_{17} s_{19} s_{9} - s_{10} s_{15} s_{16} s_{20} s_{9} - 
  2 s_{10} s_{12} s_{19} s_{20} s_{9} \nn \\ && + s_{15} s_{18} s_{20} s_{5} s_{9} - s_{17} s_{19} s_{20} s_{5} s_{9} + 
  s_{16} s_{20}^2 s_{5} s_{9} + s_{15} s_{17} s_{20} s_{9}^2 + s_{12} s_{20}^2 s_{9}^2)\,.
\eea
The knowledge of $h_1$ allows us to repeat the Euclidean algorithm. We reduce the polynomials \eqref{eq:QRSLoci} by \eqref{eq:h1} and again 
obtain the second common factor from the residues of all polynomials reading
\bea \label{eq:h2}
h_2&=& s_{10}^2 s_{19} (s_{15} s_{16} + s_{12} s_{19})  - s_{10} \big[ s_{15}^2 s_{18} s_{9} + s_{19} (-s_{17} s_{19} s_{5} + s_{16} s_{20} s_{5} + 2 s_{12} s_{20} s_{9})\nn \\  && + s_{15} (-s_{18} s_{19} s_{5} + s_{17} s_{19} s_{9} + s_{16} s_{20} s_{9})\big] + s_{20} \big[s_{18} s_{5} (-s_{19} s_{5} + s_{15} s_{9}) \nn\\ && + s_{9} (-s_{17} s_{19} s_{5} + s_{16} s_{20} s_{5} + s_{15} s_{17} s_{9} + s_{12} s_{20} s_{9})\big]\,.  
\eea
To confirm that these polynomials define the codimension two locus we were looking for, we check that 
all the constraints \eqref{eq:QRSLoci} are in the ideal generated by $\langle h_1, h_2 \rangle$.

Finally, if there are no more smaller ideals, i.e.~special solutions, of $g_3^m=\hat{g}_4^m=0$  we expect its remaining solutions 
to be generic and to support matter charged under only the section $\hat{s}_m$, i.e.~matter with charges $q_m=1$, and 
$q_n=0$ for $n\neq m$. In summary, we find that matter at a generic point of the following loci has the following charges, 
\beq \label{eq:LocusWSF}
\text{
\begin{tabular}{|c|c|c|c|} \hline
Generic point in locus & $q_Q$ & $q_R$ & $q_S$ \\ \hline
$g_{2}^{QR}=g_{3}^Q=0$ \rule{0pt}{14pt} & $1$ & $1$ & $0$ \\ 
$(g_{2}^{QS})^\prime=(g_{3}^S)^\prime=0$ \rule{0pt}{14pt} & $1$ & $0$ & $1$ \\ 
$(g_{2}^{RS})^\prime=(g_{3}^S)^\prime=0$ \rule{0pt}{14pt}& $1$ & $0$ & $1$ \\  
$h_1=h_2=0$ & $1$ & $1$ & $1$ \\ \hline
$g_3^Q=\hat{g}_{4}^Q=0$ \rule{0pt}{14pt} &  1 & 0 & 0 \\ 
$g_3^R=\hat{g}_{4}^R=0$ \rule{0pt}{14pt} &  $0$ & 1 & 0 \\ 
$g_3^S=\hat{g}_{4}^S=0$ \rule{0pt}{14pt} &  $0$ & 0 & 1 \\ \hline
\end{tabular}
}
\eeq
In each of these six cases we checked explicitly the factorization of  the complete intersection \eqref{eq:DegenerationOfQuadrics} 
for $\mathcal{E}$ into an $I_2$-curve, then computed the intersections of the sections $\hat{s}_P$, $\hat{s}_m$, $m=Q,\,R,\,S$
and obtained the charges by applying the charge formula \eqref{eq:U1charge}.

\subsection{6D Matter Muliplicities and Anomaly Cancellation}
\label{sec:MatterMultiplicities}

In this section we specialize to six-dimensional F-theory compactifications on an
elliptically fibered Calabi-Yau threefolds $\hat{X}_3$ over a general two-dimensional base $B$ with generic 
elliptic fiber given by \eqref{eq:EonBlP3}. We work out the spectrum of charged hypermultiplets, that transform
in the 14 different singlet representations found in sections \ref{sec:singsOfRationalSections} and \ref{sec:SingsWSF}.
To this end, we compute the explicit expressions for the multiplicities of these 14 hypermultiplets. 
We show consistency of this charged spectrum by checking anomaly-freedom. 

The matter multiplicities are given by the homology class of the irreducible locus that supports a given matter representation. 
As discussed above, some of these irreducible matter loci can only be expressed as prime ideals, of which we can not directly compute the 
homology classes.
Thus, we have to compute matter multiplicities successively, starting from  the complete intersections 
$\text{Loc}_{\text{CI}}$ in \eqref{eq:LocusWSF} that support multiple matter fields of different type. We found, that at the generic point
of the complete intersection $\text{Loc}_{\text{CI}}$ one type of matter is supported, but at special points $\text{Loc}_{s}^i$ different matter
fields are located. We summarize this as 
\beq \label{eq:subLoci}
	\cup_i\text{Loc}_s^i\subset \text{Loc}_{\text{CI}}\,.
\eeq
Thus,
first we calculate all multiplicities of  matter located at all these special loci $\text{Loc}_s^i$ and then subtract them from the complete intersection 
$\text{Loc}_{\text{CI}}$ in which they are contained  with a certain degree. 
This degree is given by the order of vanishing of resultant, that has already been used in a similar context in \cite{Cvetic:2013nia}. It is defined as 
follows. Given two polynomials 
$(r,s)$ in the variables $(x,y)$, if $(0,0)$ is a zero of both polynomials, its degree is given by the order of vanishing of the resultant 
$h(y):=\text{Res}_x(r,s)$ at $y=0$.

This is a straightforward calculation when the variables $(x,y)$ are pairs of the coefficients $s_i$. However, for more complicated loci we will need 
to treat full polynomials $(p_1,p_2)$ as these variables, for example $x=\tilde{s}_7$, $y=\tilde{s}_9$ or $x=\delta g_6$, $y=g_9$. In this case we have to solve 
for two coefficients $s_i$, $s_j$ from $\{p_1 = x,p_2=y\}$, then replace them in $(r,s)$ and finally proceed to take the resultant in $x$ and 
$y$. 

There is one technical caveat, when we are considering polynomials $(p_1,p_2)$ that contain multiple different matter multiplets. 
We choose the coefficients $s_i$, $s_j$ in such a way that  the variables $(x,y)$ 
only parametrize the locus of the hypermultiplets we are interested in. This is achieved by choosing $s_i$, $s_j$ we are solving for so that 
the polynomials  of the locus we are \textit{not} interested in appear as denominators and are, thus, forbidden.
For example, let us look at the loci $|M_3^Q|=|M_3^P|=0$. This complete intersection contains the loci of the hypermultiplets
with charges $(0,0,2)$ at the generic point and with charges $(0,1,2)$ at the special locus $s_9=s_{19}=0$, c.f.~\eqref{eq:TwoMinorsCharges}, 
respectively, \eqref{eq:ThreeMinorCharges}. Let us focus on the former hypermultiplets. We set
\beq
|M_3^Q|=s_{18} s_9 - s_{19} s_8\equiv x\,,  \qquad |M_3^P|=s_{10} s_{19} - s_{20} s_9\equiv y\,,
\eeq
and solve for $s_8$ and $s_{20}$ to obtain
\beq
s_8 = \frac{(s_{18} s_9 - x)}{s_{19}}\,, \qquad s_{20} = \frac{(s_{10} s_{19} + y)}{s_9}\,.
\eeq
From this, it is clear the locus $s_9=s_{19}=0$ corresponding to  hypermultiplets with charges $(0,1,2)$ is excluded because of the denominators.
Thus, $(x,y)$ indeed parametrize the locus of the hypermultiplets of charges $(0,0,2)$.

We begin the computation of multiplicities with the simplest singularities in \ref{sec:TwoMinors} located at the vanishing-loci of two coefficients 
$s_i=s_j=0$. Their multiplicities are directly given by their homology classes, that are simply the product of the classes $[s_i]$, $[s_j]$. We 
obtain
\beq \label{eq:MultpTwoMinors}
\text{
\begin{tabular}{|c|c|c|c|c|c|} \hline
Loci &  $q_Q$ & $q_R$ & $q_S$ & Multiplicity\rule{0pt}{14pt} \\ \hline
$s_{8}=s_{18}=0$ & 1 & $1$ & $-1$ & $[s_8]\cdot[s_{18}] \rule{0pt}{14pt} $ \\ 
$s_{9}=s_{19}=0$ &  0 & $1$ & $2$ & $[s_9]\cdot[s_{19}]$ \rule{0pt}{14pt} \\
$s_{10}=s_{20}=0$ &  1 & 0 & $2$ & $[s_{10}]\cdot[s_{20}]$\rule{0pt}{14pt}  \\ \hline
\end{tabular}
}
\eeq

Next we proceed to calculate the multiplicities of the loci given by the vanishing of three minors given in  \eqref{eq:ThreeMinorCharges}. The most 
direct way of obtaining these multiplicities is by using the Porteous formula to obtain the first Chern class of a determinantal variety. However,  we 
will use here a simpler approach that yields the same results. 

It was noted in section \ref{sec:ThreeMinors}, that the locus described by the vanishing of the three minors can be 
equivalently represented as the vanishing of only  two minors, after excluding the zero locus from the vanishing of the 
two coefficients $s_i$, $s_j$ that appear in both two minors. Thus, the multiplicities can be calculated by multiplying 
the homology classes of the two minors and subtracting the homology class $[s_i]\cdot[s_j]$ of the locus 
$s_i=s_j=0$. 

For example the multiplicity of the locus $|M_3^Q|=|M_2^Q|=|M_1^Q|=0$ can be obtained from multiplying the 
classes of $|M_3^Q|=|M_1^Q|=0$ and subtracting the multiplicity of the locus $s_8=s_{18}=0$ that satisfies these two 
equations, but not $M_2^Q=-s_6 s_{19}+s_9s_{16}$: 
\bea \label{eq:x0-11}
x_{(-1,0,1)}\!\!&\!=\!\!&\! [|M_3^Q|] \cdot [|M_1^Q|]-[s_8]\cdot[s_{18}]  \\
\!\!&\!=\!\!&\!([p_2]^b)^2\!+ [p_2]^b \!\cdot( \hat{\cS}_7+\cdot \tilde{\cS}_7- 3  \tilde{\cS}_9) + [K_B^{-1}] \cdot \tilde{\cS}_7  + \tilde{\cS}_7^2  - \hat{\cS}_7 \cdot \cS_9 -  2 \tilde{\cS}_7 \cdot \cS_9 + 2 \cS_9^2\,, \nn
\eea
Here we denote the multiplicity of hypermultiplets with charge $(q_Q,q_R,q_S)$ by $x_{(q_Q,q_R,q_S)}$, indicate 
homology classes of sections of line  bundles by $[\cdot]$, as before, and employ
\eqref{eq:sectionsFibration}, \eqref{eq:sP3ToP2I} and the  divisors defined in \eqref{eq:S7S9divisors}
to obtain the second line. Calculating the other multiplicities in a similarly we obtain
\beq \label{eq:DetVarMulties}
\text{
\begin{tabular}{|c|c|c|} \hline
Charges & Loci  & Multiplicity \rule{0pt}{14pt} \\ \hline
$(-1,0,1)$ & $|M_3^Q|=|M_2^Q|=|M_1^Q|=0$ & $x_{(-1,0,1)}=[|M_1^Q|]\cdot[|M_3^Q|]-[s_8]\cdot[s_{18}] $ \rule{0pt}{14pt} \\ 
$(0,-1,1)$ & $|M_3^R|=|M_2^R|=|M_1^R|=0$  & $x_{(0,-1,1)}=[|M_1^R|]\cdot[|M_3^R|]-[s_8]\cdot[s_{18}] $ \rule{0pt}{14pt} \\ 
$(-1,-1,-2)$ & $|M_3^P|=|M_2^P|=|M_1^P|=0$  & $x_{(-1,-1,-2)}=[|M_2^P|]\cdot[|M_3^P|]-[s_{10}]\cdot[s_{20}]$ \rule{0pt}{14pt} \\
$(0,0,2)$ &$|M_3^P|=|M_3^Q|=|M_3^R|=0$  & $x_{(0,0,2)}=[|M_3^Q|]\cdot[|M_3^P|] -[s_{19}][s_9] $ \rule{0pt}{14pt} \\ \hline
\end{tabular}
}
\eeq
It is straightforward but a bit lengthy to use \eqref{eq:sectionsFibration} in combination with \eqref{eq:sP3ToP2I}, 
\eqref{eq:sP3ToP2II} to obtain, as demonstrated in \eqref{eq:x0-11}, the expressions for the 
multiplicities of all these matter fields explicitly.
We have shown one possible way of calculating the multiplicities in \eqref{eq:DetVarMulties}, 
i.e.~choosing one particular pair of minors. We emphasize that the same results for the multiplicities can be obtained 
by picking any other the possible pairs of minors.

Finally we calculate the hypermultiplets of the matter found in the WSF, as discussed in section \ref{sec:SingsWSF}. In 
each case, in order to calculate the multiplicity of the matter located at a generic point of the polynomials 
\eqref{eq:LocusWSF} we need to first identify  all the loci, which solve one particular constraint in 
\eqref{eq:LocusWSF}, but  support other charged hypermultiplets. Then, we have to find the respective orders of 
vanishing of the polynomial in \eqref{eq:LocusWSF} at these special loci using the resultant technique explained
below \eqref{eq:subLoci}. Finally, we compute the homology class of the complete intersection under consideration
in \eqref{eq:LocusWSF} subtract the homology classes of the special loci with their appropriate orders.

We start with the matter with charges $(1,1,1)$ in \eqref{eq:LocusWSF} which is located at a generic point of the locus 
$h_1=h_2=0$. In this case, the degree of vanishing of the other loci are given by
\beq \label{eq:111subloci}
\text{
\begin{tabular}{|c||c|c|c|c|c|c|c|} \hline
Charge &  $\!\!x_{(1,1,-1)}\!\!$ & $\!\!x_{(0,1,2)}\!\!$ & $\!\!x_{(1,0,2)}\!\!$& $\!\!x_{(-1,0,1)}\!\!$ &$\!\!x_{(0,-1,1)}\!\!$ &$\!\!x_{(-1,-1,-2)}\!\!$&$\!\!x_{(0,0,2)}\!\!$ \rule{0pt}{14pt} \\ \hline
$(1,1,1)$ &  0 & 1 & 1 & 0 & 0 & 4 & 0 \rule{0pt}{14pt} \\ \hline
\end{tabular}
}
\eeq
Here we labeled the loci that are contained in $h_1=h_2=0$ by the multiplicity  of matter which supported on them. 
We note that the other six matter fields in \eqref{eq:LocusWSF} do not appear in this table, because the matter with 
charges $(1,1,1)$ is contained in their loci, as we demonstrate next.
This implies that the multiplicity of the hypermultiplets with charge $(1,1,1)$ is given by
\bea \label{eq:mult111}
x_{(1,1,1)} &=& [h_1] \cdot [h_2]- x_{(0,1,2)}-x_{(1,0,2)}-4 x_{(-1,-1,-2)}\,,\nn \\ 
 &=& 4 [K_B^{-1}]^2 - 3 ([p_2]^b)^2 - 2 [K_B^{-1}] \hat{\cS}_7 - 3 ([p_2]^b) \cdot\hat{\cS}_7 - 2 [K_B^{-1}]\cdot \tilde{\cS}_7 - 3 ([p_2]^b) \cdot\tilde{\cS}_7 \nn \\ && - 
 2 \hat{\cS}_7\cdot \tilde{\cS}_7 + 2 [K_B^{-1}] \cS_9 + 9 ([p_2]^b) \cS_9 + 5 \hat{\cS}_7 \cdot\cS_9 + 5 \tilde{\cS}_7 \cdot\cS_9 - 8 \cS_9^2 \,,
\eea
where the first term is the class of the complete intersection $h_1=h_2=0$ and the three following terms
are the necessary subtractions that follow from \eqref{eq:111subloci}.
The homology classes of $h_1$, $h_2$ can be obtained by determining the 
class of one term in \eqref{eq:h1}, respectively, \eqref{eq:h2} using 
\eqref{eq:sectionsFibration}.

Proceeding in a similar way for the hympermultiplets with charges $(1,0,1)$, $(0,1,1)$ and $(1,1,0)$ we get the following orders of vanishing of the loci supporting the remaining matter fields:
\beq \label{eq:110subloci}
\text{
\begin{tabular}{|c||c|c|c|c|c|c|c|c|} \hline
Charges& $\!\!x_{(1,1,-1)}\!\!$ & $\!\!x_{(0,1,2)}\!\!$ & $\!\!x_{(1,0,2)}\!\!$& $\!\!x_{(-1,0,1)}\!\!$ &$\!\!x_{(0,-1,1)}\!\!$ &$\!\!x_{(-1,-1,-2)}\!\!$&$\!\!x_{(0,0,2)}\!\!$ & $\!\!x_{(1,1,1)}\!\!$ \rule{0pt}{14pt} \\ \hline
$(1,0,1)$&   0 & 0 & 4 & 0 & 0 & 4 & 0 & 1 \rule{0pt}{14pt} \\ 
$(0,1,1)$&   0 & 4 & 0 & 0 & 0 & 4 & 0 & 1 \rule{0pt}{14pt} \\ 
$(1,1,0)$ &   1 & 0 & 0 & 0 & 0 & 1 & 0 & 1\rule{0pt}{14pt} \\ \hline
\end{tabular}
}
\eeq
We finally obtain the multiplicities of these matter fields by computing the homology class of 
the corresponding complete intersection in \eqref{eq:LocusWSF} and subtracting the multiplicities the matter fields
contained in these complete intersections with the degrees determined in \eqref{eq:110subloci}.
We obtain
\bea \label{eq:mult101}
x_{(1,0,1)} &=& 2 [K_B^{-1}]^2 + 3 ([p_2]^b)^2 + 2 [K_B^{-1}] \hat{\cS}_7 + 3 ([p_2]^b) \hat{\cS}_7 - 3 [K_B^{-1}] \tilde{\cS}_7 + 3 ([p_2]^b) \tilde{\cS}_7 \nn \\ && + 
 2 \hat{\cS}_7 \tilde{\cS}_7 + \tilde{\cS}_7^2 + 2 [K_B^{-1}] \cS_9 - 9 ([p_2]^b) \cS_9 - 5 \hat{\cS}_7 \cS_9 - 4 \tilde{\cS}_7 \cS_9 + 6 \cS_9^2 \,, \nn \\
 x_{(0,1,1)} &=& 2 [K_B^{-1}]^2 + 3 ([p_2]^b)^2 - 3 [K_B^{-1}] \hat{\cS}_7 + 3 ([p_2]^b) \hat{\cS}_7  + \hat{\cS}_7^2 + 2 [K_B^{-1}] \tilde{\cS}_7 \nn \\ && + 
 3 ([p_2]^b) \tilde{\cS}_7 + 2 \hat{\cS}_7 \tilde{\cS}_7 + 2 [K_B^{-1}] \cS_9 - 9 ([p_2]^b) \cS_9 - 4 \hat{\cS}_7 \cS_9 - 5 \tilde{\cS}_7 \cS_9 + 
 6 \cS_9^2\,, \nn \\
 x_{(1,1,0)} &=& 2 [K_B^{-1}]^2 + 3 ([p_2]^b)^2 + 2 [K_B^{-1}] \hat{\cS}_7 + 3 ([p_2]^b) \hat{\cS}_7 + 2 [K_B^{-1}] \tilde{\cS}_7 + 3 ([p_2]^b) \tilde{\cS}_7 \nn \\ && + 
 \hat{\cS}_7 \tilde{\cS}_7 - 3 [K_B^{-1}] \cS_9 - 9 ([p_2]^b) \cS_9 - 4 \hat{\cS}_7 \cS_9 - 4 \tilde{\cS}_7 \cS_9 + 7 \cS_9^2 \,.
\eea

Finally for the hypermultiplets of charges $(1,0,0)$, $(0,1,0)$ and $(0,0,1)$ we obtain the following degrees of vanishing of the loci supporting the other matter fields:
\beq \label{eq:100subloci}
\hspace{-0.1cm}{\footnotesize
\text{
\begin{tabular}{|c||c|c|c|c|c|c|c|c|c|c|c|} \hline
$\!\!\text{Charges}\!\!$  & $\!\!x_{(1,1,-1)}\!\!\!$ & $\!\!x_{(0,1,2)}\!\!$ & $\!\!x_{(1,0,2)}\!\!\!$& $\!\!x_{(-1,0,1)}\!\!\!$ &$\!\!x_{(0,-1,1)}\!\!\!$ &$\!\!x_{(-1,-1,-2)}\!\!\!$&$\!\!x_{(0,0,2)}\!\!\!$  & $\!\!x_{(1,0,1)}\!\!$ & $\!\!x_{(0,1,1)}\!\!\!$ & $\!\!x_{(1,1,0)}\!\!$ &  $\!\!x_{(1,1,1)}\!\!\!\!\!$ \rule{0pt}{14pt} \\ \hline
$\!\!(1,0,0)\!\!$ &  1 & 0 & 1 & 1 & 0 & 1 & 0 & 1 & 0 & 1 & 1\rule{0pt}{14pt} \\ 
$\!\!(0,1,0)\!\!$ &  1 & 1 & 0 & 0 & 1 & 1 & 0 & 0 & 1 & 1 & 1\rule{0pt}{14pt} \\ 
$\!\!(0,0,1)\!\!$& 1 & 16 & 16 & 1 & 1 & 16 & 16 & 1 & 1 & 0 & 1 \rule{0pt}{14pt} \\ \hline
\end{tabular}
}}
\eeq
\normalsize
Again we first computing the homology class of 
the complete intersection in \eqref{eq:LocusWSF} supporting the hypermultiplets with charges $(1,0,0)$, $(0,1,0)$,
respectively, $(0,0,1)$ and subtracting the multiplicities the matter fields
contained in these complete intersections with the degrees determined in \eqref{eq:100subloci}.
We obtain
\bea \label{eq:mult100}
x_{(1,0,0)} &=& 4 [K_B^{-1}]^2 - 3 ([p_2]^b)^2 - 2 [K_B^{-1}] \hat{\cS}_7 - 3 ([p_2]^b) \hat{\cS}_7 + 2 [K_B^{-1}] \tilde{\cS}_7 - 3 ([p_2]^b) \tilde{\cS}_7 \nn \\ && - 
 \hat{\cS}_7 \tilde{\cS}_7 - 2 \tilde{\cS}_7^2 - 2 [K_B^{-1}] \cS_9  + 9 ([p_2]^b) \cS_9 + 4 \hat{\cS}_7 \cS_9 + 5 \tilde{\cS}_7 \cS_9 - 6 \cS_9^2 \,, \nn \\
x_{(0,1,0)} &=& 4 [K_B^{-1}]^2 - 3 ([p_2]^b)^2 + 2 [K_B^{-1}] \hat{\cS}_7 - 3 ([p_2]^b) \hat{\cS}_7 - 2 \hat{\cS}_7^2 - 2 [K_B^{-1}] \tilde{\cS}_7 \nn \\ &&  - 
 3 ([p_2]^b) \tilde{\cS}_7 - \hat{\cS}_7 \tilde{\cS}_7 - 2 [K_B^{-1}] \cS_9 + 9 ([p_2]^b) \cS_9 + 5 \hat{\cS}_7 \cS_9 + 4 \tilde{\cS}_7 \cS_9 - 
 6 \cS_9^2 \,, \nn \\
 x_{(0,0,1)} &=& 4 [K_B^{-1}]^2 - 4 ([p_2]^b)^2 + 2 [K_B^{-1}] \hat{\cS}_7 - 4 ([p_2]^b) \hat{\cS}_7 - 2 \hat{\cS}_7^2 + 2 [K_B^{-1}] \tilde{\cS}_7 - 
 4 ([p_2]^b) \tilde{\cS}_7 \nn \\ && - 2 \hat{\cS}_7 \tilde{\cS}_7 - 2 \tilde{\cS}_7^2 + 2 [K_B^{-1}] \cS_9 + 12 ([p_2]^b) \cS_9 + 6 \hat{\cS}_7 \cS_9 + 
 6 \tilde{\cS}_7 \cS_9s - 10 \cS_9^2 \,. \nn
\eea

We conclude by showing that  the spectrum of the theory we have calculated is anomaly-free, which serves
also as a physically motivated consistency check for the completeness of analysis of codimension two
singularities presented in sections \ref{sec:singsOfRationalSections} and \ref{sec:SingsWSF}. 
We refer to \cite{Erler:1993zy,Honecker:2006dt} for a general account on anomaly cancellation and to 
\cite{Park:2011ji,Morrison:2012ei,Cvetic:2013nia} for the explicit form of the anomaly cancellation conditions adapted 
to the application to F-theory, c.f.~for example Eq. (5.1) in \cite{Cvetic:2013nia}. Indeed, we readily check
that the spectrum \eqref{eq:MultpTwoMinors}, \eqref{eq:DetVarMulties}, \eqref{eq:mult111}, \eqref{eq:mult101} and
\eqref{eq:mult100} together with  the height pairing matrix $b_{mn}$ reading
\bea 
 b_{mn} = -\pi( \sigma(\hat{s}_m) \cdotp \sigma(\hat{s}_n) ) =\left( \begin{array}{ccc}
                    -2[K_B] & -[K_B] & \cS_9-\tilde{\cS}_7  - [K_B] \\
                    -[K_B] & -2[K_B] & \cS_9-\hat{\cS}_7  - [K_B] \\
                    \cS_9-\hat{\cS}_7  - [K_B] & \cS_9-\hat{\cS}_7  - [K_B] & 2(\cS_9-[K_B])  \\
                   \end{array} \right)_{mn}\,.
\eea
with $m,n=1,2,3$ all mixed gravitational-Abelian and purely-Abelian anomalies in Eq. (5.1) of \cite{Cvetic:2013nia} are 
canceled.

\section{Conclusions}
\label{sec:Conclusions}

In this work we have analyzed F-theory compactifications with
U$(1)\times$U$(1)\times$U(1) gauge symmetry that are obtained by 
compactification on the most general
elliptically fibered Calabi-Yau manifolds with a rank three Mordell-Weil 
group. We have found that the natural presentation of 
the resolved elliptic fibration with three rational sections 
is given by a Calabi-Yau complete intersection $\hat{X}$ with general 
elliptic fiber given by the unique Calabi-Yau 
complete intersection in $\text{Bl}_3\mathbb{P}^3$. We have shown that all
F-theory vacua obtained by compactifying on a generic $\hat{X}$ over 
a given general base $B$ are classified by certain reflexive polytopes
related to the nef-partition of $\text{Bl}_3\mathbb{P}^3$.

We have analyzed the geometry of these elliptically fibered Calabi-Yau 
manifolds $\hat{X}$ in detail, in particular the singularities of the
elliptic fibration at codimension two in the base $B$. This way we could
identify the 14 different matter representations of F-theory 
compactifications on $\hat{X}$ to four and six dimensions. 
We have found three
matter representations that are simultaneously charged 
under all three U(1)-factors, most notably  a tri-fundamental representation. 
This unexpected representation is present because of the presence
of a codimension two locus in $B$, along which all the four constraints 
in \eqref{eq:QRSLoci}, 
$\delta g_2^{QR}$, $\delta g_2^{QS}$, $g_3^Q$ and $\hat{g}_4^Q$, 
miraculously vanish simultaneously. We could explicitly identify the two 
polynomials describing this codimension two locus algebraically in 
\eqref{eq:h1}, \eqref{eq:h2} by application of the Euclidean
algorithm. These results point to  an intriguing structure of codimension two singularities encoded in the elliptic fibrations with higher rank  Mordell-Weil groups.

We also determined the multiplicities of the massless charged 
hypermultiplets in six-dimensional F-theory
compactifications with general two-dimensional base $B$. The key 
to this analysis was the identification of the codimension two loci of 
all matter fields, which required a two-step strategy where
first the singularities of the rational sections in the resolved 
fibration with $\text{Bl}_3\mathbb{P}^3$-elliptic fiber have to be 
determined and then the remaining singularities that are visible in the 
singular Weierstrass form. We note that the loci of the former
matter are determinantal varieties, whose homology classes we determine
in general.
The completeness of our strategy has been cross-checked by verifying 6D
anomaly cancellation.

We would like to emphasize certain technical aspects in the analysis of 
the elliptic fibration. Specifically, we constructed three birational maps 
of the elliptic
curve $\mathcal{E}$ in $\text{Bl}_3\mathbb{P}^3$ to three different
elliptic curves in $dP_2$. On the level of the toric ambient 
spaces $\text{Bl}_3\mathbb{P}^3$ and $dP_2$ these maps are toric 
morphisms. The general elliptic curves in these toric varieties are 
isomorphic, whereas the map breaks down for the degenerations of 
$\mathcal{E}$ in section \ref{sec:TwoMinors}. Besides loop-holes
of this kind, we expect the degeneration of 
$\text{Bl}_3\mathbb{P}^3$-elliptic fibrations to be largely captured
by the degenerations of the non-generic $dP_2$-fibrations.

It would be important for future works to systematically  add non-Abelian 
gauge groups to the rank three Abelian sector of F-theory
on $\hat{X}$. This requires to classify the possible ways to engineer
appropriate codimension one singularities of the elliptic fibration
of $\hat{X}$. A straightforward way to obtain many explicit constructions of 
non-Abelian gauge groups  is to employ the aforementioned
birational maps to $dP_2$, because every codimension  one singularity of 
the $dP_2$-elliptic fibration  automatically induces an according 
singularity of the 
$\text{Bl}_3\mathbb{P}^3$-elliptic fibration. In particular, many concrete
$I_4$-singularities, i.e.~SU(5) groups, can be obtained by application
of the constructions of $I_4$-singularities of $dP_2$-elliptic fibrations 
in \cite{Borchmann:2013jwa,Cvetic:2013nia,Braun:2013yti}. However, it 
would be important to analyze whether all codimension one singularities 
of $\hat{X}$ are induced by singularities of the corresponding 
$dP_2$-elliptic fibrations.
For phenomenological applications, it would then be relevant to determine
the matter representations for all 
possible SU(5)-GUT sectors that can be realized in Calabi-Yau manifolds 
$\hat{X}$ with $\text{Bl}_3\mathbb{P}^3$-elliptic fiber. 
Compactifications with $\text{Bl}_3\mathbb{P}^3$-elliptic fiber might lead to new implications for 
for particle physics: e.g.,  the appearance of 
${\mathbf{10}}$-representations  with different U(1)-factors, which does not  seem to 
appear in the rank-two Mordell-Weil constructions,  and  the intriguing possibility 
for the appearance of 
$\mathbf{5}$-representations charged under all three U(1)-factors,
i.e.~quadruple-fundamental representations,
which  are not present in perturbative  Type II compactifications.

Furthermore, for explicit 4D GUT-model building, it would be 
necessary to combine the analysis of this work with the techniques of 
\cite{Cvetic:2013uta} to obtain chiral four-dimensional compactifications
of F-theory. The determination of chiral indices of 4D matter requires the 
determination of all matter surfaces as well as the construction of the 
general $G_4$-flux on Calabi-Yau fourfolds $\hat{X}$ with general elliptic 
fiber in $\text{Bl}_3\mathbb{P}^3$, most desirable in the presence of an 
interesting GUT-sector. Furthermore the structure of Yukawa couplings  should 
to be determined by an analysis of codimension three singularities 
of the fibration.

\subsubsection*{Acknowledgments}

We gratefully acknowledge discussions and correspondence with Lasha 
Berezhiani,Yi-Zen Chu, Thomas Grimm and in particular Antonella Grassi and 
Albrecht Klemm. M.C.  thanks CERN Theory Division  and Aspen Center for Physics for hospitality. D.K. thanks the Bethe Center for Theoretical Physics Bonn and the Mitchell Institute at Texas A\&M University for hospitality.  The research is supported by the  DOE grant DE-SC0007901 (M.C., H.P., D.K.), the NSF String Vacuum Project Grant No. NSF PHY05-51164 (H.P.), Dean's Funds for Faculty Working Group (M.C., D.K.),
the Fay R. and Eugene L.Langberg Endowed Chair (M.C.) and the Slovenian Research Agency
(ARRS) (M.C.).

\appendix
\section{The Weierstrass Form of the Elliptic Curve with Three Rational Points}
\label{app:WSFdirectly}

The main text made extensive use of the mapping of the elliptic curve $\mathcal{E}$ with Mordell-Weil rank three to the 
Calabi-Yau hypersurface in $dP_2$. Specifically, the calculation of the coordinates of the rational points, the 
Weierstrass form and the discriminant were all performed employing the results for the $dP_2$-elliptic curve in 
\cite{Cvetic:2013nia}. Following \cite{Morrison:2012ei,Cvetic:2013nia}, that we refer when needed, in this appendix we 
calculate the Weierstrass form and the coordinates of the three ratinal points directly from the three elliptic curve 
$\mathcal{E}$.

In order to motivate the approach below, we briefly summarize how to obtain the Tate form of an elliptic curve with the 
zero point $P$. Given an 
elliptic curve  with one marked point $P$, we 
can obtain the Tate equation with respect to this point by finding the sections of $\mathcal{O}(kP)$, $k=1,...,6$. The coordinate $z$ will be the only section of $\mathcal{O}(P)$, the coordinate $x$ is a section of $\mathcal{O}(2P)$ independent of $z^2$, and $y$ is a section of $\mathcal{O}(3P)$ independent of $z^3$ and $xz$. The Tate equation is obtained from the linear relation between the sections of $\mathcal{O}(6P)$. 

\subsubsection*{Coordinates $x$, $y$ and $z$}

To obtain the birational map from the complete intersection \eqref{eq:1st} in $\mathbb{P}^3$ to the Tate form, we need 
to construct the Weierstrass coordinates $x$, $y$ and $z$ as sections of the line bundles $\mathcal{O}(kP)$ on 
$\mathcal{E}$ with $k=1,2,3$. In section \ref{sec:EasQuadricInt} we found a basis for the bundle 
$\mathcal{M}=\mathcal{O}(P+Q+R+S)$, as well as a basis for $\mathcal{M}^2$ and a choice of  basis for $\mathcal{M}^3$. 
The sections of $\mathcal{O}(kP)$ are obtained from linear combinations of $\mathcal{O}(kM)$ that vanish with degree 
$k$ at the points $Q$, $R$ and $S$.

From the discussion in section \ref{sec:EasQuadricInt}, the section $z$ can be taken to be $z:=u'$. To find $x$, 
we take 
an eight-dimensional basis of $H^0(\mathcal{E},\mathcal{M}^2)$ and construct the most general linear combination. The 
coefficient of $u'^2$ is set to zero in order for $x$ to be independent of $z^2$. Thus, the ansatz for the variable $x$ 
reduces to
\begin{equation} \label{eq:XTate}
x:=at'^2+cv'^2+dw'^2+et'u'+fu'v'+gu'w'+hv'w'\,.
\end{equation}
Six out of the seven coefficients are fixed by imposing zeroes of order two at the three points $Q$, $R$ and $S$. The 
last coefficient can be eliminated by an overall scaling. Solving the constraints but  keeping $h$ as the overall 
scaling coefficient, we obtain
\bea
& a = \frac{h (s_{10} s_{19} - s_{20} s_{9})^2 }{(s_{10} s_{18} - s_{20} s_{8}) (-s_{19} s_{8} + s_{18} s_{9})}\,, \qquad c = d =0  \,, \qquad f = h\frac{ (s_{19} s_{6} - s_{16} s_{9})}{s_{19} s_{8} - s_{18} s_{9}}\,, \qquad g = h\frac{ (s_{10} s_{17} - s_{20} s_{7})}{s_{10} s_{18} - s_{20} s_{8}}\,, & \nn 
\eea
\bea
 e&=&- h   s_{10}\frac{  s_{18} s_{19} s_{5} - s_{19} s_{20} s_{6} + s_{19}^2 s_{7} + s_{15} s_{19} s_{8} - 
        2 s_{15} s_{18} s_{9} - s_{17} s_{19} s_{9} - s_{16} s_{20} s_{9}}{(s_{10} s_{18} - s_{20} s_{8}) (-s_{19} s_{8} + s_{18} s_{9})}   \nn \\   
    && -h\frac{ 
     s_{10}^2 s_{16} s_{19}  + s_{20} \big[s_{9} (s_{18} s_{5} + s_{20} s_{6} + s_{15} s_{8} + s_{17} s_{9}) - 
        s_{19} (2 s_{5} s_{8} + s_{7} s_{9})\big]}{(s_{10} s_{18} - s_{20} s_{8}) (-s_{19} s_{8} + s_{18} s_{9})} \,.\nn  
\eea
Finally consider $y \in \mathcal{O}(3P)$ as a section linearly independent of $u^3$ and $u x$. We make the ansatz
\begin{equation} \label{eq:YTate}
y:=\tilde{a}t'^3+\tilde{c}v'^3+\tilde{d}w'^3+\tilde{f}t'u'^2+\tilde{g}u'^2v'+\tilde{h}u'^2w'+\tilde{i}u'v'^2
+\tilde{j}u'w'^2+\tilde{k}u'v'w'+\tilde{l}v'^2w'\,,
\end{equation}
where again, all but one of the coefficients can be fixed by demanding $y$ to have zeroes of degree three at $Q$, $R$ and $S$ and the free coefficient is an overall scaling. The solutions of these coefficients are long and not illuminating, thus we will not be presented here but can be provided on request.

\subsubsection*{Tate equations and Weierstrass form}

Once the sections $x$, $y$ and $z$ are known, we impose the Tate form
\beq \label{eq:Tate}
 y^2 + a_1 y x z + a_3 y z^3 = x^3 + a_4 x^2 z^4 + a_6 z^6 \,
\eeq
to hold in the ideal generated by the complete intersection \eqref{eq:1st}.  
First we exploit the free scalings of $x$ and $y$ to obtain coefficients equal to 
one in front of the monomials $x^3$ and $y^2$ in  \eqref{eq:XTate} and \eqref{eq:YTate}. 
Then we compute all the monomials in equation \eqref{eq:Tate} after inserting $z=u'$, \eqref{eq:XTate} and 
\eqref{eq:YTate} and reduce by the ideal generated by the polynomials \eqref{eq:1st}.
Finally, from a comparison of coefficient, we obtain 23 equations that can be solved uniquely for the five 
Tate coefficients $a_i$. Unfortunately the results are long and not illuminating and are again provided on request. 

From the Tate form \eqref{eq:Tate}, the Weierstrass form
\beq \label{eq:WSF1}
		y^2=x^3+fxz^4+gz^6
\eeq
is obtained by the variable transformation
\beq \label{eq:varTrafoToWSF}
	x\mapsto  x+\tfrac{1}{12}b_2 z^2\,,\qquad y\mapsto y+\tfrac12 a_1 x z+\tfrac12 a_3z^3
\eeq
with the following definitions
\bea\label{eq:defsWSF}
	f&=&-\tfrac{1}{48}(b_2^2-24 b_4)\,,\qquad g=-\tfrac{1}{864}(-b_2^3 + 36 b_2 b_4 -216 b_6)\,,
	\nn\\
	b_2&=&a_1^2+4a_2\,,\qquad b_4=a_1 a_3+2 a_4\,,
	\qquad b_6=a_3^2+4 a_6\,,\nn\\
	\Delta&=&-16(4f^3+27 g^2)= -8b_4^3+\tfrac{1}{4}b_2^2b_4^2+9b_2 b_4 b_6-
	\tfrac{1}{4}b_6 b_2^3 - 27 b_6 \,.
\eea

\subsubsection*{Rational points in the Weierstrass form}

Equipped with the Weierstrass form \eqref{eq:WSF1} of the curve, we calculate the coordinates 
$[x^m:y^m:z^m]=[g^m_2:g^m_3:b^m]$ of all the rational points $m=P,Q,R,S$. By construction the point $P$ is mapped to 
the zero section, that is the point $[\lambda^2:\lambda^3:0]$. 

The coordinates of the other points are all obtained through the following procedure: Let us call the generic point $N$ 
with Tate coordinates $[x^N:y^N:z^N]$. First, we find a section of degree two, denoted $x^\prime$,  that vanishes with 
degree three at the point $N$. In this case we need to make use of the full basis of $\mathcal{O}(2M)$ that includes 
$u^2$. The vanishing at degree two already fixes most of the coefficients as in \eqref{eq:XTate}. The condition of 
vanishing at degree three fixes the new coefficient of $u^2$. Restoring the variables $x$ and $z$ we obtain 
\beq
 x^\prime |_N =  x+ \tilde{g}_m z^2\,.
\eeq
Then, the coordinate $x^N$ of $N$ is given in terms of $z^N$ by requiring $x^\prime|_N=0$. The coordinate $y^N$ is 
determined by inserting the values for $z^N$, $x^N$ into the Tate form \eqref{eq:Tate}. Finally, the coordinates
in Weierstrass form are obtained by the transformations \eqref{eq:varTrafoToWSF}.

We summarize our results for the coordinates of the rational points $Q$, $R$ and $S$ in the following. We obtain the 
coordinates of the form
\bea \label{eq:pointsWeierstrass}
\left[x^Q,y^Q,z^Q\right] &=& [g_2^Q : g_3^Q : 1]\,, \\
\left[ x^R, y^R,z^R \right] &=& [g_2^R: g_3^R : 1]\,, \\
\left[x^S,y^S,z^S \right] &=& [g_2^S:g_3^S:(s_{10} s_{19} - s_{20} s_{9})] \,,
\eea
where we have made the following definitions:
\bea \label{eq:xyQ}
g_2^Q&=&\frac{1}{12} \Big[ 8 (s_{10} s_{15} - s_{20} s_{5}) (s_{18} s_{6} - s_{16} s_{8}) + (s_{10} s_{16} + s_{18} s_{5} - 
     s_{20} s_{6} + s_{19} s_{7} - s_{15} s_{8} - s_{17} s_{9})^2 \nn  \\ && - 
   4 (s_{10} s_{12} - s_{2} s_{20} + s_{17} s_{5} - s_{15} s_{7}) (s_{19} s_{8} - s_{18} s_{9}) \nn \\&&  + 
   4 (s_{18} s_{2} + s_{17} s_{6} - s_{16} s_{7} - s_{12} s_{8}) (s_{10} s_{19} - s_{20} s_{9})\Big]\,,\\
g_3^Q&=&\frac{1}{2} \Big[(-s_{10} s_{15} + s_{20} s_{5}) (-s_{18} s_{6} + s_{16} s_{8}) (-s_{10} s_{16} - s_{18} s_{5} + 
      s_{20} s_{6} - s_{19} s_{7} + s_{15} s_{8} + s_{17} s_{9}) \nn\\ && - (s_{10} s_{15} - 
      s_{20} s_{5}) (s_{18} s_{2} + s_{17} s_{6} - s_{16} s_{7} - s_{12} s_{8}) (-s_{19} s_{8} + 
      s_{18} s_{9}) \nn \\&&- (s_{10} s_{12} - s_{2} s_{20} + s_{17} s_{5} - s_{15} s_{7}) (s_{18} s_{6} - 
      s_{16} s_{8}) (s_{10} s_{19} - s_{20} s_{9}) \nn \\&&+ (s_{17} s_{2} - s_{12} s_{7}) (s_{19} s_{8} - 
      s_{18} s_{9}) (s_{10} s_{19} - s_{20} s_{9})\Big]\,,
\eea
\bea \label{eq:xyR}
g_2^R &=& \frac{1}{12} \Big[ -4 (s_{10} s_{18} - s_{20} s_{8}) (s_{19} s_{2} - s_{16} s_{5} + s_{15} s_{6} - s_{12} s_{9}) \nn \\ && + 
   8 (-s_{18} s_{7} + s_{17} s_{8}) (s_{19} s_{5} - s_{15} s_{9}) + (s_{10} s_{16} - s_{18} s_{5} - 
     s_{20} s_{6} + s_{19} s_{7} + s_{15} s_{8} - s_{17} s_{9})^2  \nn \\ && - 
   4 (s_{18} s_{2} - s_{17} s_{6} + s_{16} s_{7} - s_{12} s_{8}) (s_{10} s_{19} - s_{20} s_{9})\Big]\,,\\
g_3^R&=&\frac{1}{2} \Big[ (s_{18} s_{2} - s_{17} s_{6} + s_{16} s_{7} - s_{12} s_{8}) (s_{10} s_{18} - s_{20} s_{8}) (s_{19} s_{5} - 
      s_{15} s_{9})  \nn \\ && + (s_{18} s_{7} - s_{17} s_{8}) (s_{19} s_{5} - s_{15} s_{9}) (-s_{10} s_{16} + 
      s_{18} s_{5} + s_{20} s_{6} - s_{19} s_{7} - s_{15} s_{8} + s_{17} s_{9})  \nn \\ && + (s_{16} s_{2} - 
      s_{12} s_{6}) (s_{10} s_{18} - s_{20} s_{8}) (s_{10} s_{19} - s_{20} s_{9})  \nn \\ && - (s_{18} s_{7} - 
      s_{17} s_{8}) (s_{19} s_{2} - s_{16} s_{5} + s_{15} s_{6} - s_{12} s_{9}) (s_{10} s_{19} - s_{20} s_{9})\Big]\,,
\eea
\bea \label{eq:xyS}
g_2^S &=&\frac{1}{12} \Big\lbrace 12 (s_{10} s_{18} - s_{20} s_{8})^2 (s_{19} s_{5} - s_{15} s_{9})^2  \nn \\ && + (s_{10} s_{19} - 
      s_{20} s_{9})^2 \big[8 (-s_{18} s_{7} + s_{17} s_{8}) (s_{19} s_{5} - s_{15} s_{9})  \nn \\ &&+ (s_{10} s_{16} - 
        s_{18} s_{5} - s_{20} s_{6} + s_{19} s_{7} + s_{15} s_{8} - s_{17} s_{9})^2  \nn \\ && - 
      4 (s_{18} s_{2} - s_{17} s_{6} + s_{16} s_{7} - s_{12} s_{8}) (s_{10} s_{19} - s_{20} s_{9})\big]  \nn \\ && + 
   4 (s_{10} s_{18} - s_{20} s_{8}) (-s_{10} s_{19} + 
      s_{20} s_{9}) \times \big[ \nn \\  &&  -3 (s_{19} s_{5} - s_{15} s_{9}) (s_{10} s_{16} - s_{18} s_{5} - s_{20} s_{6} + 
         s_{19} s_{7} + s_{15} s_{8} - s_{17} s_{9})  \nn \\ && + 
      2 (s_{19} s_{2} - s_{16} s_{5} + s_{15} s_{6} - s_{12} s_{9}) (-s_{10} s_{19} + s_{20} s_{9})\big] \Big\rbrace\,,
     \eea
      \bea 
g_3^S&=&\frac{1}{2} \Big\lbrace 2 (s_{10} s_{18} - s_{20} s_{8})^3 (s_{19} s_{5} - s_{15} s_{9})^3 \\ \nn && + (s_{10} s_{18} - 
      s_{20} s_{8}) (s_{10} s_{19} - 
      s_{20} s_{9})^2 \big[ 2 (-s_{18} s_{7} + s_{17} s_{8}) (s_{19} s_{5} - s_{15} s_{9})^2 \\ \nn && + (s_{19} s_{5} - 
         s_{15} s_{9}) (s_{10} s_{16} - s_{18} s_{5} - s_{20} s_{6} + s_{19} s_{7} + s_{15} s_{8} - 
         s_{17} s_{9})^2 \\ \nn && - (s_{18} s_{2} - s_{17} s_{6} + s_{16} s_{7} - s_{12} s_{8}) (s_{19} s_{5} - 
         s_{15} s_{9}) (s_{10} s_{19} - s_{20} s_{9}) \\ \nn && + (s_{19} s_{2} - s_{16} s_{5} + s_{15} s_{6} - 
         s_{12} s_{9}) (s_{10} s_{16} - s_{18} s_{5} - s_{20} s_{6} + s_{19} s_{7} + s_{15} s_{8} - 
         s_{17} s_{9}) (s_{10} s_{19} - s_{20} s_{9}) \\ \nn && + (-s_{16} s_{2} + s_{12} s_{6}) (s_{10} s_{19} - 
         s_{20} s_{9})^2\big] \\ \nn && + (-s_{18} s_{7} + s_{17} s_{8}) (-s_{10} s_{19} + 
      s_{20} s_{9})^3 \times  \\ \nn && \big[ -(s_{19} s_{5} - s_{15} s_{9}) (s_{10} s_{16} - s_{18} s_{5} - s_{20} s_{6} + 
         s_{19} s_{7} + s_{15} s_{8} - s_{17} s_{9}) \nn \\ && + (s_{19} s_{2} - s_{16} s_{5} + s_{15} s_{6} - 
         s_{12} s_{9}) (-s_{10} s_{19} + s_{20} s_{9})\big] \nn \\ && + (s_{10} s_{18} - s_{20} s_{8})^2 (s_{19} s_{5} -
       s_{15} s_{9}) (-s_{10} s_{19} + 
      s_{20} s_{9}) \times \nn \\ && \big[ -3 (s_{19} s_{5} - s_{15} s_{9}) (s_{10} s_{16} - s_{18} s_{5} - s_{20} s_{6} + 
         s_{19} s_{7} + s_{15} s_{8} - s_{17} s_{9}) \nn \\ && + 
      2 (s_{19} s_{2} - s_{16} s_{5} + s_{15} s_{6} - s_{12} s_{9}) (-s_{10} s_{19} + s_{20} s_{9})\big] \Big\rbrace.
\eea

\section{Nef-partitions}
\label{app:nefparts}

Here we recall the very basic definitions and results about nef-Partitions. We refer for example to 
\cite{cox1999mirror} for a detailed mathematical account.

\textbf{Definition} Let $X=\mathbb{P}_{\nabla}$ be a toric variety with a corresponding polytope $\nabla$, a normal fan of the polytope $\nabla$ and rays $\rho\in\Sigma(1)$ with associated divisors $D_{\rho}$. Given a partition of $\Sigma(1)=I_1 \cup \cdots \cup I_k$, into $k$ disjoint subsets, there are divisors $E_j= \sum _{\rho\in I_j} D_\rho$ such that $-K_X=E_1+\cdots+E_k$. This decomposition is called a nef-partition if for each $j$, $E_j$ is a a Cartier divisor spanned by its global sections.

We denote the convex hull of the rays in $I_j$ as $\nabla_j$ and their dual polytopes by $\Delta_j$, which are defined as
\beq \label{eq:dualNEF}
	\Delta_j=\{m\in \mathbb{Z}^3| \langle m,\rho_i \rangle\geq-\delta_{ij}\,\text{ for } \rho_i\in \nabla_j\}.
\eeq
The generic global sections, $h_j$ of $D_j$ are computed according to the expression
\beq \label{eq:sectionNEFpart}
	h_j=\sum_{m\in \Delta_j\cap \mathbb{Z}^3}a_{m}\prod_{j=1}^k\prod_{\rho_i\in \nabla_j}x_i^{\langle m,\rho^i\rangle+\delta_{ij}}\,,\qquad \delta_{ij}=\left\{\begin{matrix}
	1\,\,\text{ for }  \rho_i\in \nabla_j\\
	0\,\, \text{ else.\phantom{.........}}
\end{matrix}\right.	 
\eeq

\bibliographystyle{utphys}	
\bibliography{ref}

\end{document}